\documentclass[runningheads]{llncs}
\usepackage{graphicx}
\usepackage{subcaption}

\usepackage{tikz}
\usepackage{comment}
\usepackage{amsmath,amssymb}
\usepackage{color}
\usepackage{multirow}

\usepackage{ulem}
\newcommand{\QxQ}{Q$\times$Q}
\usepackage{algorithm} 
\usepackage{algpseudocode} 
\usepackage{placeins}

\usepackage[accsupp]{axessibility}

\begin{document}
\pagestyle{headings}
\mainmatter

\title{PyNET-Q$\times$Q: An Efficient PyNET Variant for Q$\times$Q Bayer Pattern Demosaicing in CMOS Image Sensors} % Replace with your title

\titlerunning{PyNET-\QxQ{}}
\author{Minhyeok Cho\inst{1} \and
Haechang Lee\inst{2} \and
Hyunwoo Je\inst{2} \and
Kijeong Kim\inst{2}\and
Dongil Ryu\inst{2} \and
Albert No\inst{1}}
\authorrunning{Cho et al.}
\institute{Hongik University, Seoul 04066, Korea \\
\email{albertno@hongik.ac.kr}, \email{mincho@mail.hongik.ac.kr} \and
SK hynix, Icheon 17336, Korea\\
\email{\{haechang.lee,hyunwoo.je,kijeong1.kim,dongil.ryu\}@sk.com}}
\maketitle

\begin{abstract}

Deep learning-based image signal processor (ISP) models for mobile cameras can generate high-quality images 
that rival those of professional DSLR cameras.
However, their computational demands often make them unsuitable for mobile settings.
Additionally, modern mobile cameras employ non-Bayer color filter arrays (CFA) such as
Quad Bayer, Nona Bayer, and Q$\times$Q Bayer to enhance image quality,
yet most existing deep learning-based ISP (or demosaicing) models focus primarily on standard Bayer CFAs.
In this study, we present PyNET-Q$\times$Q, a lightweight demosaicing model specifically designed
for Q$\times$Q Bayer CFA patterns,
which is derived from the original PyNET.
We also propose a knowledge distillation method called progressive distillation 
to train the reduced network more effectively.
Consequently, PyNET-Q$\times$Q contains less than 2.5\% of the parameters of the original PyNET 
while preserving its performance.
Experiments using Q$\times$Q images captured by a prototype Q$\times$Q camera sensor 
show that PyNET-Q$\times$Q outperforms existing conventional algorithms in terms of texture and edge reconstruction,
despite its significantly reduced parameter count.
Code and partial datasets can be found at \url{https://github.com/Minhyeok01/PyNET-QxQ}.

\end{abstract}

\section{Introduction}

\begin{figure}
\begin{subfigure}{0.48\textwidth}
\includegraphics[width=0.95\textwidth]{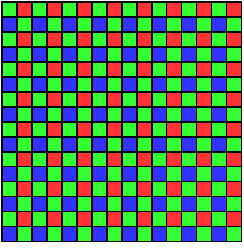}% 
\caption{Standard Bayer CFA}
\label{fig:bayercfa}
\end{subfigure}
\hfill
\begin{subfigure}{0.48\textwidth}
\includegraphics[width=0.95\textwidth]{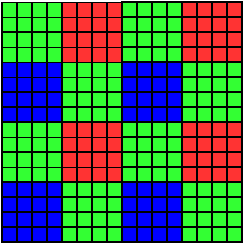}
\caption{Q$\times$Q Bayer CFA}
\label{fig:qxqcfa}
\end{subfigure}
\caption{Standard Bayer and Q$\times$Q  Bayer color filter arrays.}
\label{fig:CFA}
\end{figure}

As the demand for higher-quality images continues to grow,
mobile camera sensors are becoming more integrated, leading to smaller pixel sizes.
However, these smaller pixel sizes directly affect image quality, especially in low-light conditions.
Pixel-binning has been proposed to address this issue by grouping nearby pixels to create larger effective pixel sizes,
which improves the signal-to-noise ratio (SNR) in low-light settings while maintaining high-resolution images 
under bright conditions.
Examples of this technique include Quad and Nona color filter arrays (CFA),
used in recent flagship smartphones such as the Samsung Galaxy S21 Ultra and Xiaomi Mi 11 Ultra.
The Q$\times$Q Bayer CFA (Figure~\ref{fig:qxqcfa}, which groups $4\times4$ pixels, has also been proposed.

Camera sensors capture RAW images as single-channel images since each pixel can only record single-color information.
Image signal processors (ISPs) then convert these RAW images into high-quality RGB images 
through processes such as demosaicing, denoising, white balancing, and gamma correction.
Although deep learning-based ISP techniques have greatly 
improved image reconstruction quality~\cite{pynet,eednet,delnet,csanet},
most existing methods focus on standard Bayer CFAs (Figure~\ref{fig:bayercfa}), 
with only a few considering non-Bayer CFAs~\cite{nona,2021demosic}.
With non-Bayer CFAs causing different image statistics compared to standard Bayer filtered images,
ISPs should be re-optimized, particularly for the latest Q$\times$Q Bayer CFAs.
However, no ISPs specifically designed for Q$\times$Q Bayer CFA currently exist.

Another challenge for mobile cameras is the limited computational resources available.
The trend of over-parameterization in deep learning models~\cite{transfomersr,imagetransformer} 
is not suitable for resource-constrained environments.
PyNET~\cite{pynet}, for instance, achieves impressive performance in RAW to RGB reconstruction 
but has a parameter count of 47.55 million and a trained parameter size of 181.6 MB.

In this paper, we propose PyNET-Q$\times$Q, a next-generation deep learning-based ISP model 
explicitly designed for resource-constrained environments and Q$\times$Q Bayer CFAs.
We primarily focus on demosaicing, as other tasks like whitening and gamma correction can be subjectively 
and tunably adjusted at the software level.
PyNET-Q$\times$Q is based on the recently proposed PyNET~\cite{pynet} but is more mobile-friendly 
by significantly reducing the number of model parameters.
In addition, to address the discontinuity issue inherent in Q$\times$Q images (due to $4\times4$ grouping),
we introduce two additional techniques:
1) introducing skip connection with a gray image and 2) applying sub-pixel convolution~\cite{ESPCN,deconvolution}.

However, the reduced size of the proposed model results in lower reconstructed image quality compared 
to the original PyNET.
To address this, we incorporate a knowledge distillation strategy to improve output image quality.
Knowledge distillation~\cite{kd} is a process that transfers a distilled soft label from a larger model (teacher) 
to a smaller model (student).
We propose {\it progressive distillation} for generative models, 
which transfers knowledge from different levels of teachers,
allowing the student model to learn from the appropriate teacher level.

% Note that the similar idea is proposed in classification setup~\cite{softoutput, rezagholizadeh2021pro}.
% We train PyNET-Q$\times$Q  with progressive distillation and verify the effectiveness in generative tasks via experiments.

Note that obtaining ground truth images while training deep learning-based ISP models is challenging.
PyNET~\cite{pynet} is trained with RAW images captured by a Huawei P20 camera phone (as input) 
and images captured by a Canon 5D Mark IV DSLR camera (as ground truth output).
However, such datasets can have alignment issues and may depend on specific ISP systems with subjective components.
Most other works~\cite{demosaic2,demosaic4,demosaic5} apply Bayer CFA to common images to obtain input images,
but these filtered images have different statistics from sensor-level images due to ISP processing.
In this work, we train our model with a hybrid dataset, 
consisting of RAW 3CCD images captured by a Hitachi HV-F203SCL and the common dataset DIV2K~\cite{div2k}.
Unlike standard cameras, each pixel sensor of a 3CCD camera captures all sensor-level RGB color information,
making it more suitable as ground truth.
The corresponding Q$\times$Q input is a filtered RAW 3CCD image by Q$\times$Q Bayer CFA.
Therefore, the hybrid dataset includes a diverse range of scenes from DIV2K and more accurate color information 
from 3CCD images.
We also test the proposed model on Q$\times$Q input images obtained 
by an actual Q$\times$Q sensor (currently under development).

Our main contributions can be summarized as follows:
\begin{itemize}
\item  We propose the {\bf first mobile-friendly demosaicing model for Q$\times$Q Bayer CFA}, PyNET-Q$\times$Q.
    Specifically, we adapt PyNET for Q$\times$Q inputs by introducing skip connection with a gray image 
    and sub-pixel convolution, and then compress the model (from 181 MB to 5 MB) for mobile environments.
\item  We explore a {\bf progressive distillation strategy in generative tasks},
    which allows for more effective training of the compressed network.
\item  We incorporate {\bf sensor-level RAW 3CCD images} into the training dataset,
    making it more suitable for demosaicing tasks than common datasets.
\item  We demonstrate the performance of our model using {\bf actual Q$\times$Q input images}
    obtained from a Q$\times$Q camera sensor.
\end{itemize}

\begin{figure*}[!ht]
\centering
\begin{subfigure}{0.49\textwidth}
\includegraphics[width=0.95\textwidth]{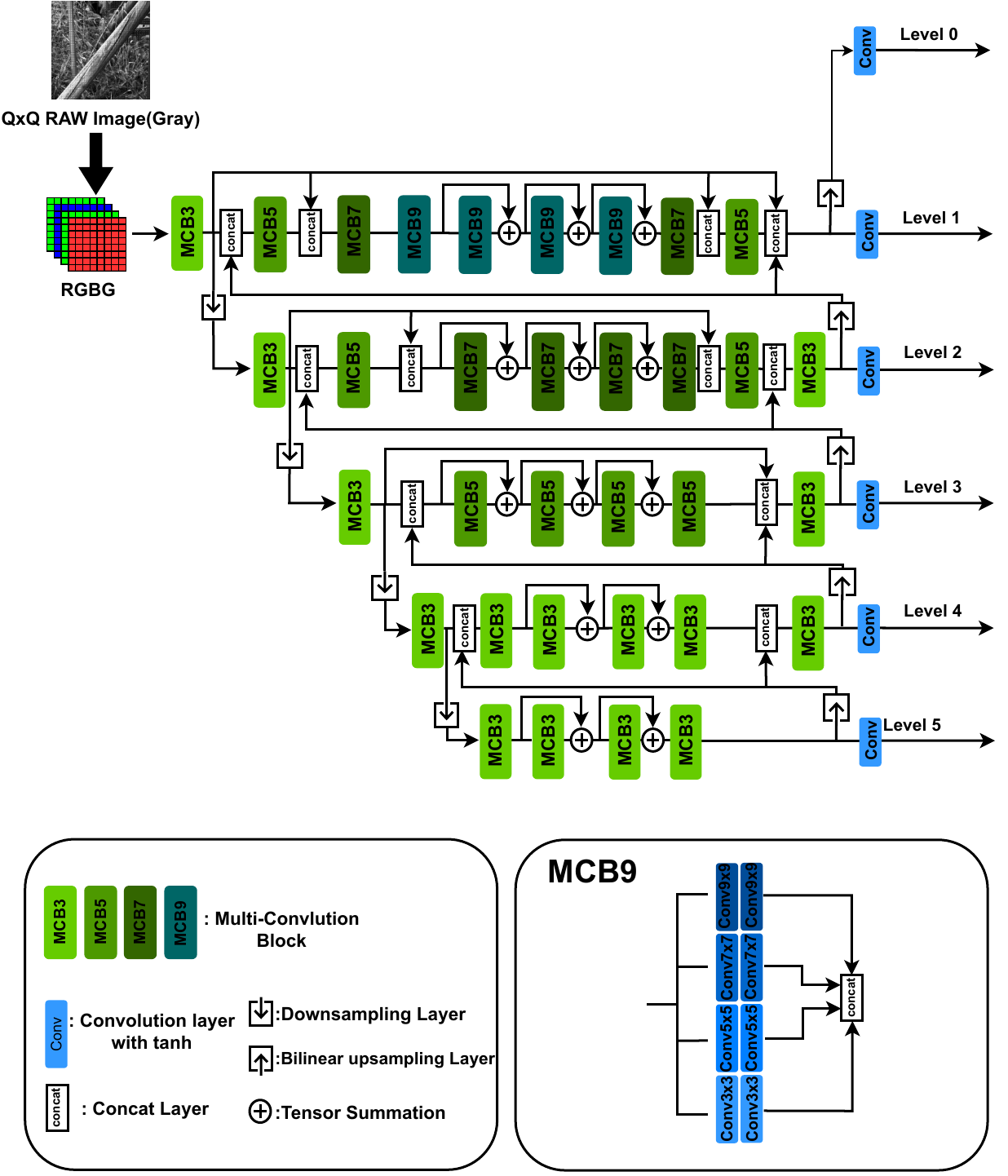}
\caption{Original PyNET}
\label{fig:orig_pynet}
\end{subfigure}
\begin{subfigure}{0.49\textwidth}
\includegraphics[width=0.95\textwidth]{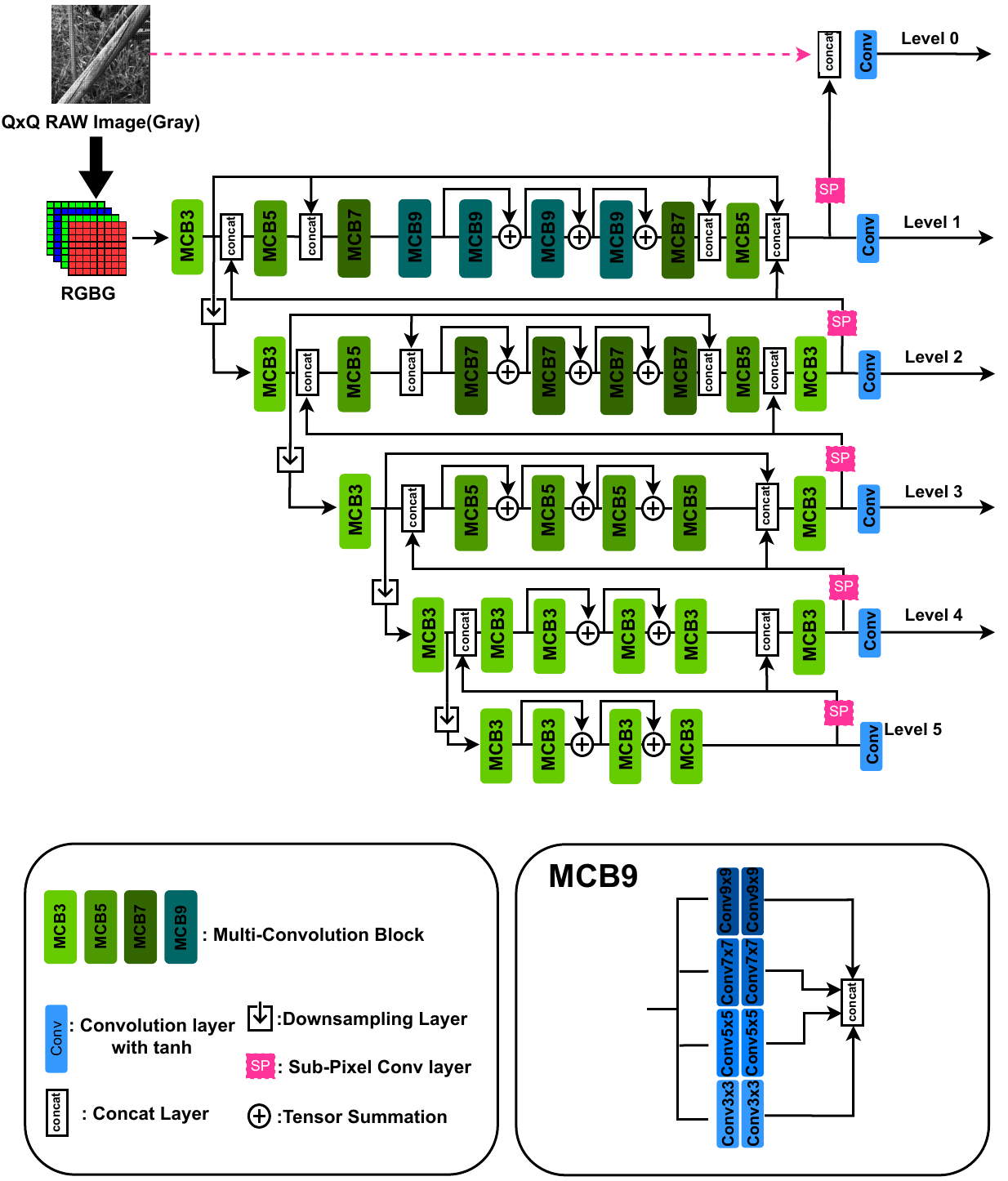}
\caption{Enhanced PyNET}
\label{fig:pynet}
\end{subfigure}
\caption{The architecture of the original and enhanced PyNET.
Enhancements (sub-pixel convolution, and skip connection with a gray image) are highlighted in magenta.}
\end{figure*}

\section{RELATED WORKS}
\subsection{Deep Learning based Demosaicing}
Deep learning-based demosaicing models have emerged due to the impressive success of deep neural networks 
across numerous computer vision tasks~ \cite{demosaic1,demosaic2,demosaic3,demosaic4,demosaic5,kokkinos2018deep}.
However, most of these models have primarily focused on Bayer CFA.
For non-Bayer CFA demosaicing, Syu et al. introduced DMCNN~\cite{demosaic7}, based on SRCNN~\cite{SRCNN},
demonstrating its effectiveness on various non-Bayer CFA inputs, including diagonal stripe~\cite{diag}, GYGM, 
and Hirakawa~\cite{hirakawa}. 
Kim et al.~\cite{2020demosaic} proposed an efficient end-to-end demosaicing model comprised of 
two pyramid networks for Quad CFA ($2\times2$), while Sharif et al.~\cite{demosaic6} suggested 
a joint demosaicing and denoising scheme utilizing depth and spatial attention.
For Nona CFA ($3\times3$), Sugawara et al.~\cite{nona} developed a GAN based on a spatial-asymmetric 
attention module to minimize artifacts. 
Kim et al.~\cite{2021demosic} demonstrated that a duplex pyramid network (DPN)~\cite{2020demosaic} achieves 
low visual artifacts and effective edge restoration in demosaicing images captured by a SAMSUNG CMOS sensor.

In addition to demosaicing-specific models, several deep learning-based ISP models have been proposed 
to replace traditional ISPs~\cite{eccvcomp}.
Recent work has concentrated on hierarchical structures, such as U-Net~\cite{unet}, 
to manage local and global features.
U-Net is a U-shaped network that processes fine and coarse features through contracting and expanding paths.
EEDNET~\cite{eednet} employed a U-Net structure with a channel attention residual dense block and clip-L1 loss.
W-Net~\cite{wnet} utilized a two-cascaded U-Net architecture with a channel attention module,
while CameraNet~\cite{cameranet} also employed two-cascaded U-Nets, consisting of Restore-Net and Enhance-Net.
Beyond U-Net-based models, DeepISP~\cite{deepisp} introduced a network by connecting two CNNs.
PyNET~\cite{pynet} significantly enhanced performance by using an inverted pyramidal structure 
with various convolution filters.
PyNET-CA~\cite{pynetca} incorporated a channel attention mechanism into PyNET,
further improving the reconstruction quality.

It is worth mentioning that super-resolution~\cite{srsurvey}, which aims to reconstruct 
a high-resolution (HR) image from a low-resolution (LR) image,
shares similarities with demosaicing as it fills in the missing information of an image.
Traditional approaches to super-resolution construct an HR image based on 
prior knowledge~\cite{sparsesrprior,gradientsr}.
In contrast, CNN-based~\cite{SRCNN,DnCNN,DRCNN,ESPCN,FSRCNN,IRCNN,VDSR}
and GAN-based~\cite{srgan,esrgan} super-resolution models generate photorealistic output images,
surpassing classical methods without using hand-crafted elements.

While other network models exist in computer vision,
such as vision transformer (ViT)~\cite{dosovitskiy2020vit,liang2021swinir}, 
and regression-based approaches~\cite{drn}, most of them are computationally intensive 
and not suitable for mobile environments.
Our model is based on PyNET due to its proven performance in ISP-related tasks 
and its compressible CNN-based architecture.

\begin{figure*}[!t]
\centering
\includegraphics[width=0.89\textwidth]{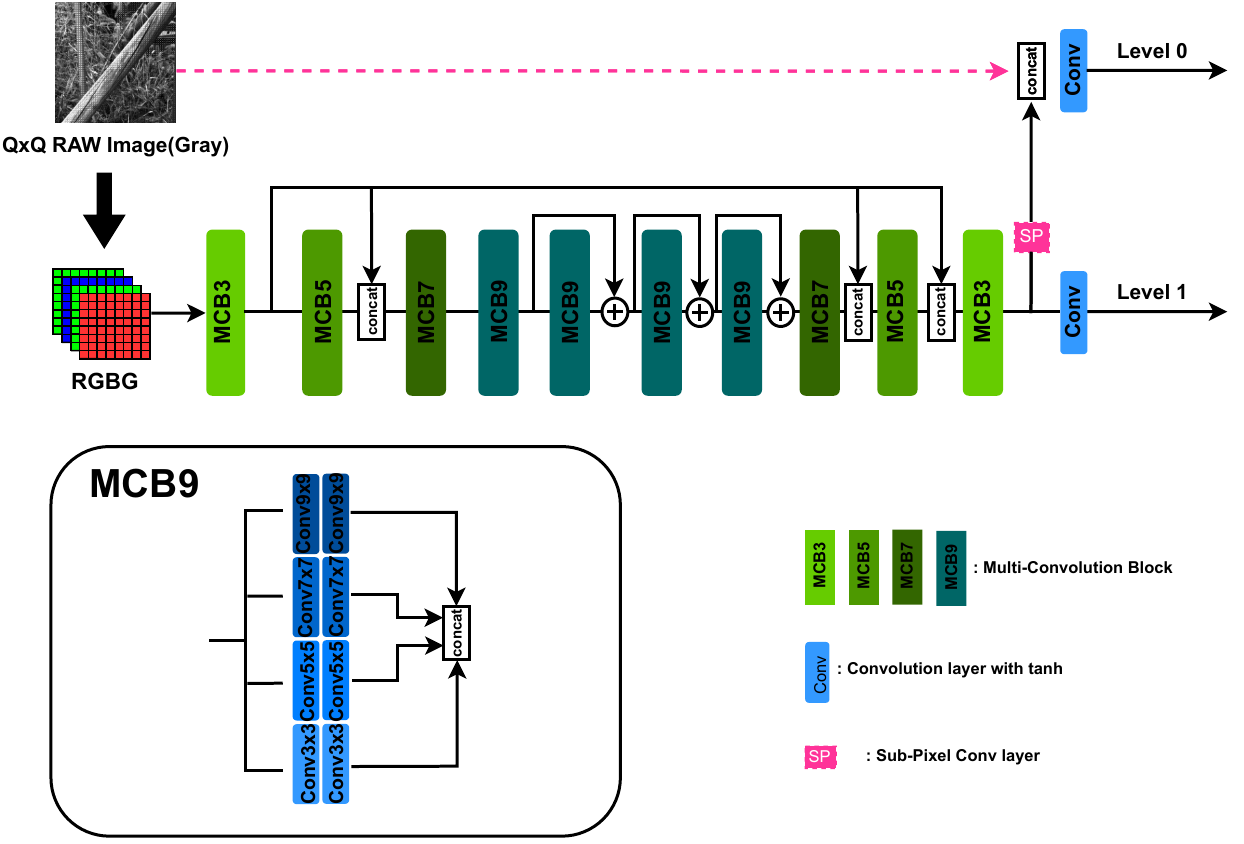}
\caption{The architecture of the proposed PyNET-Q$\times$Q .}
\label{fig:pynetqxq}
\end{figure*}

%%% PyNET
%%%%%%%%%%%%%%%%%%%%%%%%%%%%%%%%%%%%%%%%%%%%%%%%%%%%%%%%%%%%%%%%%%%%
\subsection{PyNET}
As our model structure heavily relies on PyNET~\cite{pynet},
we provide a comprehensive review of PyNET in this section.
PyNET features an inverted pyramidal structure with five levels,
where level~5 is the lowest and level~1 is the highest.
Each level works with differently scaled images; the lower levels operate on lower resolution images 
and focus on learning global features, 
while the higher levels work with higher resolution images and learn local details.
Each level of PyNET is composed of multi-convolution blocks.
The lower level contains blocks with two convolution layers that have a kernel size of $3\times3$,
whereas the higher level has parallel convolution layers with various kernel sizes
($3\times3$, $5\times5$, $7\times7$, and $9\times9$). 
Above level~1, PyNET features at level~0 that includes a convolution layer and an activation function,
which scales up the level~1 output to the target resolution.

PyNET is trained sequentially from the lowest level to the highest level.
The outputs of the lower levels are upsampled to align with the scale of features in the subsequent level.
Before applying convolution layers, the higher level concatenates the upsampled feature maps (from the previous level) and an intermediate scaled input.
All pre-trained lower levels are trained simultaneously when training the higher level.
Figure~\ref{fig:orig_pynet} depicts the network architecture of PyNET.

Despite PyNET's exceptional performance, the inverted pyramidal structure necessitates a large number of parameters,
which is not ideal for mobile environments.
We also note that the input image of PyNET comprises four channels with a half-resolution of the original RAW image,
as shown in Figure~\ref{fig:gray}.
This makes it more susceptible to discontinuity issues, particularly when the input is a Q$\times$Q image.

%%% Knowledge Distillation
%%%%%%%%%%%%%%%%%%%%%%%%%%%%%%%%%%%%%%%%%%%%%%%%%%%%%%%%%%%%%%%%%%%%
\subsection{Knowledge Distillation}
In knowledge distillation~\cite{kd}, a pre-trained, complex teacher model provides a distilled output (soft labels) 
to a smaller student model, transferring the teacher's knowledge.
Although it seems intuitive that the student model should learn better from a stronger model,
Cho and Hariharan~\cite{softoutput} demonstrated that a highly accurate teacher is not always the best choice.
They revealed that a less-trained network might be a more suitable teacher 
when the student network has limited capacity.
This suggests that the teacher's knowledge should align with the student's capabilities.
Furthermore, Rezagholizadeh et al.~\cite{rezagholizadeh2021pro} proposed progressive distillation 
for natural language processing (NLP) and classification tasks to minimize the capability gap
between the student and teacher.
Progressive distillation gradually distills knowledge from a smoother teacher to a fully-trained teacher,
allowing the student to learn from a teacher at the appropriate level.

There are various knowledge distillation methods.
Feature distillation~\cite{fitnet,at} transfers the teacher model's intermediate feature maps to a student model,
while online distillation~\cite{online0,online1,online2,online3,online4} trains both the teacher 
and student models simultaneously.
Adaptive distillation~\cite{adaptiveensemble,adaptiveteacher1} enables the student network to learn 
from multiple teachers adaptively.
Du et al.~\cite{adaptiveensemble} update the weights of knowledge distillation losses and feature losses 
during training based on teachers' gradients. 
Liu et al.~\cite{adaptiveteacher1} transfer features from multiple teachers and adaptively adjust weights 
between teachers' soft targets.

However, in generative tasks that produce an image (such as super-resolution and demosaicing),
directly applying distillation techniques is challenging since they do not have a notion of soft labels.
Consequently, many super-resolution networks distill teacher networks' features~\cite{srdistill1,fakd,lsfd} 
using an additional tool like a regressor to address dimension mismatches between the teacher and student.
Gao et al.~\cite{srdistill1} transfer the first-order statistical map (e.g., average, maximum, or minimum value)
of intermediate features, while FAKD~\cite{fakd} proposed spatial affinity of features-based distillation 
to utilize rich high-dimensional statistical information. 
LSFD~\cite{lsfd} introduced a deeper regressor comprising $3\times 3$ convolution layers to achieve 
a larger receptive field and an attention method based on the difference (between teacher and student)
that selectively focuses on vulnerable pixel locations.
PISR~\cite{pisr} added an encoder that leverages privileged information from ground truth 
and transfers knowledge through feature distillation.
Beyond super-resolution, Aguinaldo et al.~\cite{gancomp1} proposed a distillation method 
for general GANs by transferring knowledge based on the pixel-level distance between images
generated by the student and the teacher.
KDGAN~\cite{kdgan} suggested a three-player distillation with a student, a teacher, and a discriminator.
Park et al.~\cite{park2021implementation} utilized knowledge distillation for CNN-based demosaicing on FPGA.
However, most knowledge distillation approaches depend on a pre-trained fixed teacher,
overlooking the student's incapabilities during early training.
In the context of denoising tasks, HKDS~\cite{li2022multiple} shares similarities with our proposed method 
by using a two-stage distillation process with two teachers, each trained on different noise levels.
Nevertheless, our approach differs from HKDS in that we employ two teachers with varying training levels 
and it is not restricted to the denoising setup.

%%%%%%%%%%%%%%%%%%%%%%%%%%%%%%%%%%%%%%%%%%%%%%%%%%%%%%%%%%%%%%%%%%%%
%%% PyNET-\QxQ
%%%%%%%%%%%%%%%%%%%%%%%%%%%%%%%%%%%%%%%%%%%%%%%%%%%%%%%%%%%%%%%%%%%%
\section{PyNET-Q$\times$Q }

%%% Model Architecture
%%%%%%%%%%%%%%%%%%%%%%%%%%%%%%%%%%%%%%%%%%%%%%%%%%%%%%%%%%%%%%%%%%%%
\subsection{Model Architecture}
Since PyNET can handle RAW input types and boasts extraordinary performance with an easily compressible architecture,
we propose PyNET-Q$\times$Q, which is based on PyNET.
PyNET-Q$\times$Q is specifically designed for mobile devices with a lighter structure 
and additional network design tailored for Q$\times$Q input.
To create a lighter model, we retain only levels 0 and 1, removing lower levels (levels 2–5). 
Additionally, since multi-convolution blocks significantly increase FLOP counts,
PyNET-Q$\times$Q reduces the number of filters in all blocks by half.
We compensate for the model degradation resulting from compression with a distillation technique,
which we discuss in Section~\ref{subsec:progressive distillation}.

Recall that Q$\times$Q Bayer CFA groups $4\times4$ patches to obtain single-color information;
the distance between the same color patches is larger than the regular Bayer pattern input.
Therefore, Q$\times$Q input generates output images with blocking effects, 
which are further amplified when the model has reduced parameters and lower capacity.
To overcome this issue, we introduce two additional components to PyNET for Q$\times$Q images:
1) skip connection with a gray image and 2) sub-pixel convolution.
We present the network architecture of PyNET-Q$\times$Q in Figure~\ref{fig:pynetqxq}.

%%% Residual Learning
\subsubsection{Skip connection with a gray image}
The Q$\times$Q input image provides true pixel values in part,
and the corresponding pixels in the model's output should maintain consistent values.
In other words, the model should preserve information from the input Q$\times$Q RAW image 
and reconstruct the input's missing parts.
Hence, it is natural to supply an input image to the model's (near) final layer,
allowing it to focus solely on the missing part.
PyNET~\cite{pynet} has skip connections in the residual block of each level, but level~0 does not,
as it only upsamples level~1 output.

\begin{figure*}[!t]
\centering
\includegraphics[width=.9\textwidth]{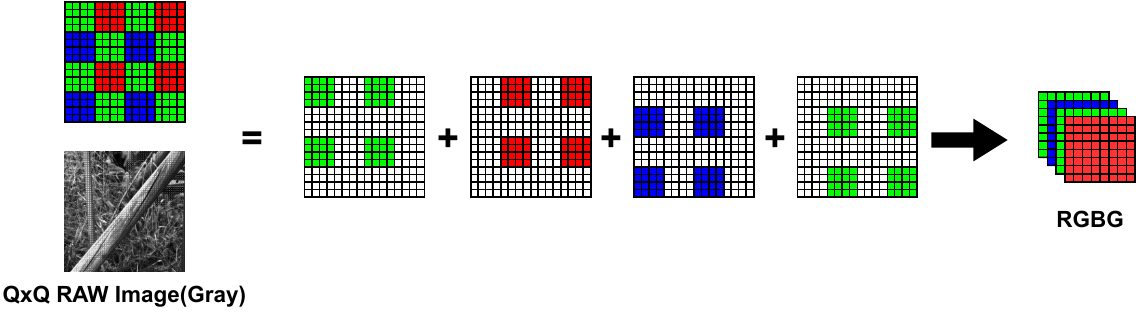}
\caption{Difference of gray image and 4 channel (RGBG) half resolution input.}
\label{fig:gray}
\end{figure*}

The proposed model includes a skip connection that provides a single-channel gray image to the final layer at level~0.
A gray image is a single-channel, full-resolution image obtained by CFA, different from a 4-channel (RGBG),
half-resolution PyNET input, as depicted in Figure~\ref{fig:gray}.
Note that there is a global skip connection in level~1,
but it does not complement the overall shape and location information.
Similar to DenseNet~\cite{huang2017densely}, the final layer at level~0 concatenates (instead of adding) 
a gray image and level~1's output.
The gray image provides a subset of the true pixel value,
which assists the next convolution layer in refining level~1's results.
The last convolution layer then produces the sensor-level RGB output image.

The gray image offers a partial ground truth for the input image, making residual learning more effective.
The effectiveness of this approach is validated by experimental results.
We provide a more detailed discussion on the gray image skip connection's effect in Section~\ref{subsec:results}.
Note that the additional skip connection only increases nine parameters per filter of the last convolution layer,
which is negligible.

%%% Pixel Shuffle 
\subsubsection{Sub-pixel convolution}
PyNET's inverted-pyramidal structure necessitates multiple upsampling steps to transform features 
from lower levels to higher levels.
PyNET~\cite{pynet} upsamples features using bilinear interpolation followed by a $3\times3$ convolution,
which results in cumulative noise in the feature and information loss due to non-invertibility.
On the other hand, deconvolution~\cite{deconvolution}, another upsampling technique with fractional stride,
exhibits a checkerboard artifact that is more critical for Q$\times$Q inputs.

Motivated by the upsampling in super-resolution~\cite{ESPCN},
we replace interpolation-based upsampling with sub-pixel convolution.
Sub-pixel convolution expands the channel by the square of the upscaling factor and rearranges it,
while deconvolution adds padding first, followed by a standard convolution.
Sub-pixel convolution is invertible and has fewer checkerboard artifacts~\cite{checkerboardfree}.
By incorporating sub-pixel convolution, we address the issues related to interpolation-based upsampling methods 
and improve the overall performance of the model for Q$\times$Q input.

%%% Progressive Distillation
%%%%%%%%%%%%%%%%%%%%%%%%%%%%%%%%%%%%%%%%%%%%%%%%%%%%%%%%%%%%%%%%%%%%
\subsection{Progressive Distillation}\label{subsec:progressive distillation}
To address the performance reduction resulting from parameter and level reduction,
we employ knowledge distillation with an enhanced PyNET as the teacher model.
The teacher model is a full-size PyNET with all five levels, the original number of filters,
and additional components for Q$\times$Q input, such as a skip connection with a gray image and sub-pixel convolution.
The {\it enhanced PyNET} teacher model is depicted in Figure~\ref{fig:pynet}, 
with the enhancements highlighted in magenta.
Taking inspiration from progressive distillation~\cite{rezagholizadeh2021pro} in classification settings,
we propose progressive distillation for generative models,
which also adaptively adjusts the teacher's level to smooth the knowledge transfer during the distillation process.

The idea behind progressive distillation for generative models is to progressively switch the teacher model 
to a more advanced one as the student becomes ready.
More specifically, let $\{T_1, ..., T_k\}$ represent the collection of teacher models 
that share the same network architecture but have been trained for a different number of epochs.
A lower index indicates a less-trained teacher, with $T_1$ being the least trained and $T_k$ being the most trained.
Initially, the student model trains independently without assistance from a teacher.
When the difference between the student's output and the ground truth plateaus, 
the student begins learning from the least-trained teacher ($T_1$).
Then, as the agreement between the student's output and the teacher's output reaches a saturation point,
the student transitions from a less-trained teacher ($T_i$) to a more-trained teacher ($T_{i+1}$).
It is important to note that $T_k$ is not a fully-trained teacher, 
as the student network's capacity is significantly smaller than the teacher's capability.
Progressive distillation enables the unprepared student to easily imitate the teacher in the early stages 
and follow the teacher's learning process by adaptively switching teachers.
In the context of classification model distillation, the distilled output can be seen as a label-smoothing regularizer.
In our case, less-trained teachers contribute to this process.

%%% Training
%%%%%%%%%%%%%%%%%%%%%%%%%%%%%%%%%%%%%%%%%%%%%%%%%%%%%%%%%%%%%%%%%%%%
\subsection{Training}
Similar to PyNET~\cite{pynet}, we sequentially train each level of PyNET-Q$\times$Q, 
starting from level~1 and moving to level~0. 
Level~1 training does not involve distillation, and the student is trained with a $2\times$ 
downscaled ground truth image.
During level~0 training, we utilize the progressive distillation scheme outlined earlier.

%%% Level 1 training 
\subsubsection{Level 1 training}
Level~1 of the proposed model is trained with $2\times$ downscaled images, without any distillation.
This is because level~1 training already provides a sufficient initialization for level~0 training 
due to the $2\times$ downscaled images. 
Let $I^{(1)}_S$ and $I^{(1)}_{GT}$ represent the demosaiced image (output of level~1) 
and the downscaled ground truth image, respectively.
Then, level~1 training minimizes the original PyNET model~\cite{pynet} loss, given by 
\begin{align}
\mathcal{L}^{(1)}_{DE} &= \left\| I^{(1)}_{S} - I^{(1)}_{GT} \right\|^2_{2} 
            + \lambda_1\cdot \mathcal{L}^{(1)}_{VGG},
\end{align}
 where $\mathcal{L}^{(1)}_{VGG}$ is a VGG~\cite{vgg}-based perceptual loss 
 that measures semantic differences of level~1 outputs with weight $\lambda_1 > 0$.

%%% Level 0 training 
\subsubsection{Level 0 training}
During level~0 training, we train the entire PyNET-Q$\times$Q model, 
initializing level~0 randomly and level~1 with the trained parameters from level~1 training.
The loss function consists of PyNET loss and distillation loss.
Let $I_S$ and $I_{GT}$ denote the demosaiced image (output of the model) and the ground truth image, respectively.
Similar to level~1 training, the PyNET loss is given by 
\begin{align}
\mathcal{L}_{DE} &= \left\| I_{S} - I_{GT} \right\|^2_{2} + \lambda_1\cdot \mathcal{L}_{VGG} + \lambda_2\cdot  \mathcal{L}_{MS-SSIM},
\end{align}
 where $\mathcal{L}_{DE}$, $\mathcal{L}_{VGG}$, and $\mathcal{L}_{MS-SSIM}$ denote demosaicing loss, 
 VGG loss, and MS-SSIM loss, respectively.
 
 For distillation loss, we distill the intermediate features of the early layers in level~1.
 This is because we removed lower levels of PyNET, and PyNET-Q$\times$Q level~1 should behave similarly 
 to the combination of all lower levels (levels 2 to 5) of PyNET.
 For feature distillation, we calculate the $L_2$ difference between intermediate features.
 However, the teacher and student have different spatial dimensions of features,
 which is why previous studies~\cite{srdistill1,fakd,lsfd} introduced additional regressors 
 or extracted statistical information from features.
 We also add $1\times 1$ regressors $R$ and compare the intermediate features of level~1. 
 For the $i$-th step of progressive distillation, when the student model $S$ learns from the $i$-th teacher $T_i$,
 the distillation loss $L^{(i)}_{DS}$ is given by 
\begin{align}
\mathcal{L}^{(i)}_{DS} &= \left\| R(F_{S}) - F_{T_i} \right\|^2_{2},
\end{align}
 where $F_S$ and $F_{T_i}$
 denote intermediate feature maps of level~1 from the student $S$ and the teacher $T_i$, respectively.

\begin{figure*}[t]
\centering
\includegraphics[width=1\textwidth]{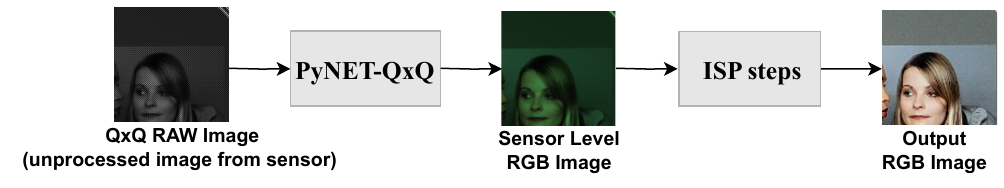}
\caption{Pipeline for evaluating Q$\times$Q demosaicing performance.
PyNET-Q$\times$Q represents our proposed algorithms, which are designed to demosaic images captured 
using a Q$\times$Q Bayer pattern.
In addition to demosaicing, the pipeline includes ISP steps that aim to enhance the quality of the resulting images 
by reducing noise, correcting color balance, adjusting contrast, and applying other techniques.}
\label{fig:sensor}
\end{figure*}

Finally, during the $i$-th step of progressive distillation,
we train PyNET-Q$\times$Q by minimizing the total loss 
\begin{align}
\mathcal{L}_{total}&= \mathcal{L}_{DE} + \alpha \cdot \mathcal{L}^{(i)}_{DS},
\end{align}
with weight $\alpha>0$.
We switch the teacher to $T_{i+1}$ if the feature difference between the student and teacher saturates.
More precisely, we consider the feature difference saturated when the variance of differences 
in the last five epochs is below a certain threshold $\sigma$.
The progressive distillation process is detailed in Algorithm~\ref{alg:progressive}.

\begin{algorithm}
\caption{Progressive distillation}\label{alg:progressive}
\textbf{Parameters:}\\ 
\hspace*{\algorithmicindent}$S$: the student model\\
\hspace*{\algorithmicindent}$\{T_1, \ldots, T_k\}$: the pre-trained teacher models
\begin{algorithmic}[1]
\State Initialize the student $S$ level 1 using the parameters from level 1 training
\State Train the student $S$ until $\|I_{S}- I_{GT}\|^2$ saturates
\For {$i=1,2,\ldots,k$}
\State Distill $T_i$ to $S$ until $\|R(F_S) -  F_{T_i}\|^2$ saturates
\EndFor
\end{algorithmic} 
\end{algorithm}

%%%%%%%%%%%%%%%%%%%%%%%%%%%%%%%%%%%%%%%%%%%%%%%%%%%%%%%%%%%%%%%%%%%%
%%% Experiments
%%%%%%%%%%%%%%%%%%%%%%%%%%%%%%%%%%%%%%%%%%%%%%%%%%%%%%%%%%%%%%%%%%%%
\section{Experiments}\label{sec:exp}
In this section, we present experimental results for PyNET-Q$\times$Q and compare them 
with traditional demosaicing techniques.
These conventional methods are not based on deep learning algorithms and consist of various hand-crafted 
interpolation algorithms, such as four-direction residual interpolation and adaptive residual interpolation.
Recall that our focus is on demosaicing, which involves reconstructing sensor-level RGB images from Q$\times$Q RAW images.
The remaining ISP steps generate the output RGB image, as illustrated in Figure~\ref{fig:sensor}.

To ensure a fair comparison, we optimize the training parameters for the base PyNET 
and apply the same hyperparameters (including the number of epochs) for all experiments.
All networks are trained using the ADAM optimizer~\cite{kingma2014adam} with $\beta_1 = 0.9$, $\beta_2 = 0.99$,
$\epsilon = 10^{-8}$, and a learning rate of $10^{-4}$.
PyNET-Q$\times$Q is trained sequentially from level~1 to level~0. 
Level~1 is optimized with $\lambda_1=0.1$, while level~0 is optimized with $\lambda_1=1$ and $\lambda_2=0.4$.
In progressive distillation, we have two teachers, $T_1$ and $T_2$,
which are trained with early stopping after the 7th and 20th epochs, respectively.
We set the weight parameter $\alpha$ to 10 to match the scale of loss values.

%%% Datasets
%%%%%%%%%%%%%%%%%%%%%%%%%%%%%%%%%%%%%%%%%%%%%%%%%%%%%%%%%%%%%%%%%%%%
\subsection{Datasets}
Most existing deep learning-based models train on high-quality, common image datasets 
such as DIV2K~\cite{div2k}, Flickr2K~\cite{flickr2k}, WED~\cite{wed}, and BSDS500~\cite{BSDS500}.
 For instance, the widely-accepted procedure in super-resolution \cite{liang2021swinir,drn,edsr,carn}
 involves downsampling the image (using a bicubic kernel) to obtain a low-resolution input image,
 while the original image serves as the high-resolution target image.
 Similarly, in demosaicing, one might want to extract an input RAW image from the dataset by applying the Bayer CFA.
 However, this approach can cause problems since high-resolution images in common datasets are outputs of an ISP system.
 Due to the nature of ISP, which includes color correction, gamma correction, and white balancing,
 the statistics of (downsampled) high-resolution images differ significantly from sensor-level inputs.
 Consequently, a model trained on downsampled data may imitate a specific ISP system.

\begin{figure*}[t]
\centering
\begin{subfigure}{0.24\textwidth}
\includegraphics[width=0.95\textwidth]{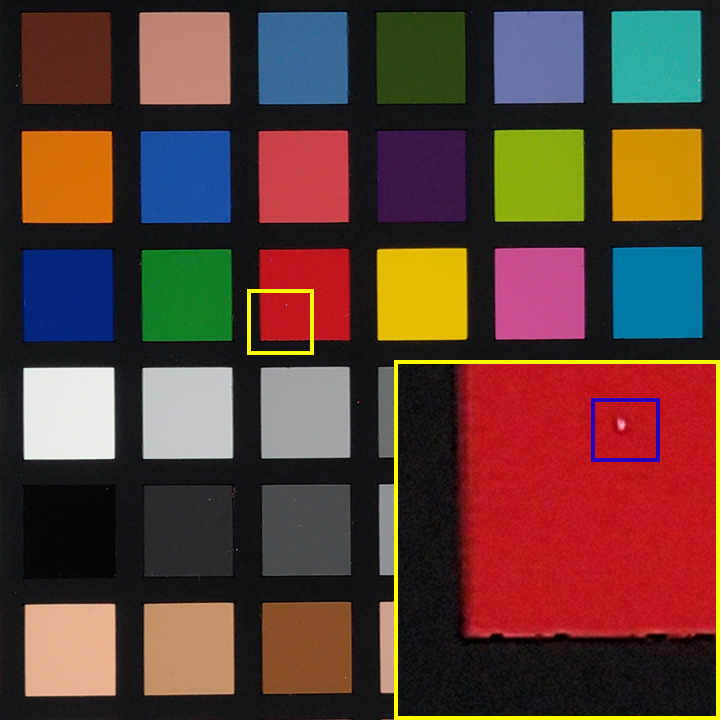}
\end{subfigure}
\hfill
\begin{subfigure}{0.24\textwidth}
\includegraphics[width=0.95\textwidth]{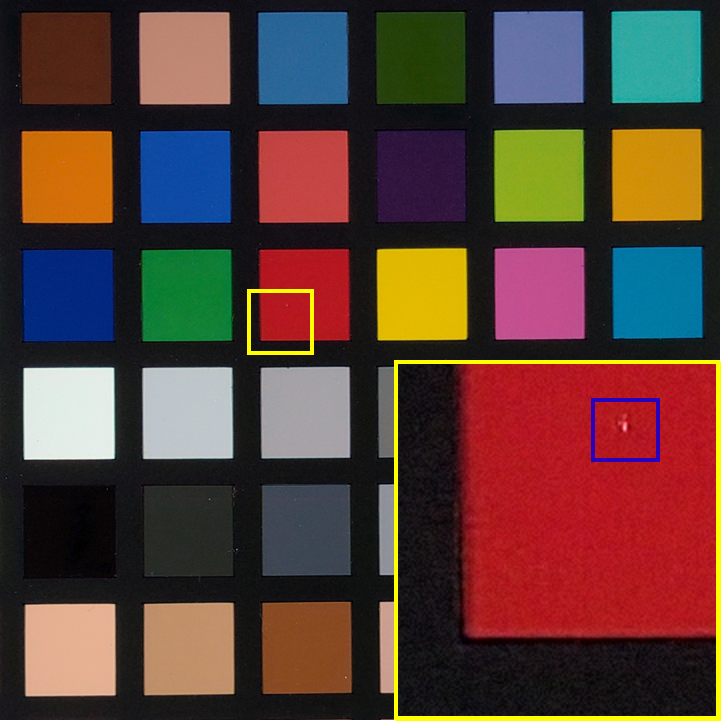}
\end{subfigure}
\hfill
\begin{subfigure}{0.24\textwidth}
\includegraphics[width=0.95\textwidth]{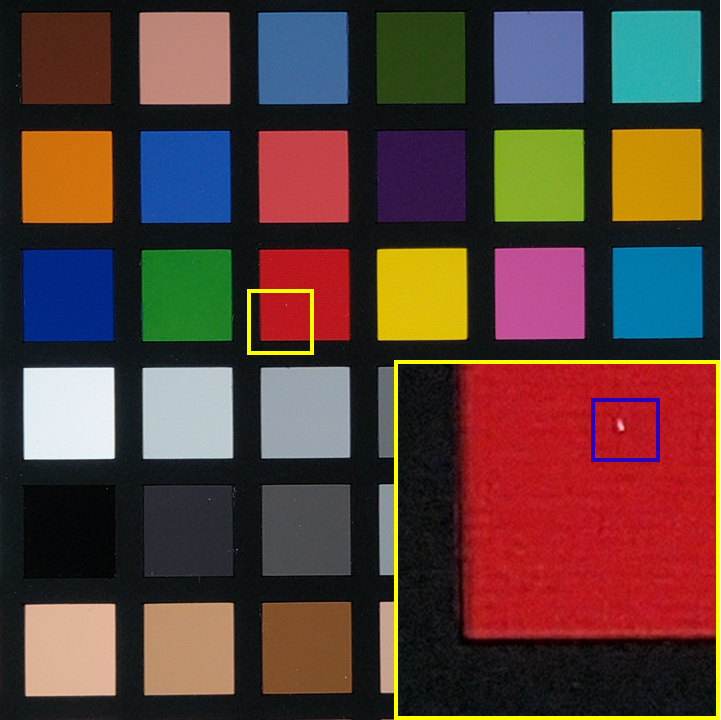}
\end{subfigure}
\hfill
\begin{subfigure}{0.24\textwidth}
\includegraphics[width=0.95\textwidth]{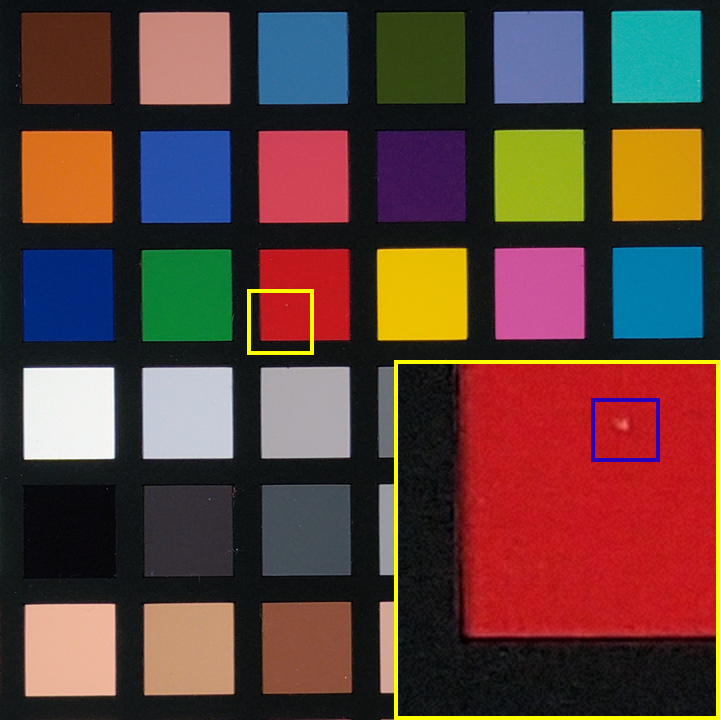}
\end{subfigure}
\\
\begin{subfigure}{0.24\textwidth}
\includegraphics[width=0.95\textwidth]{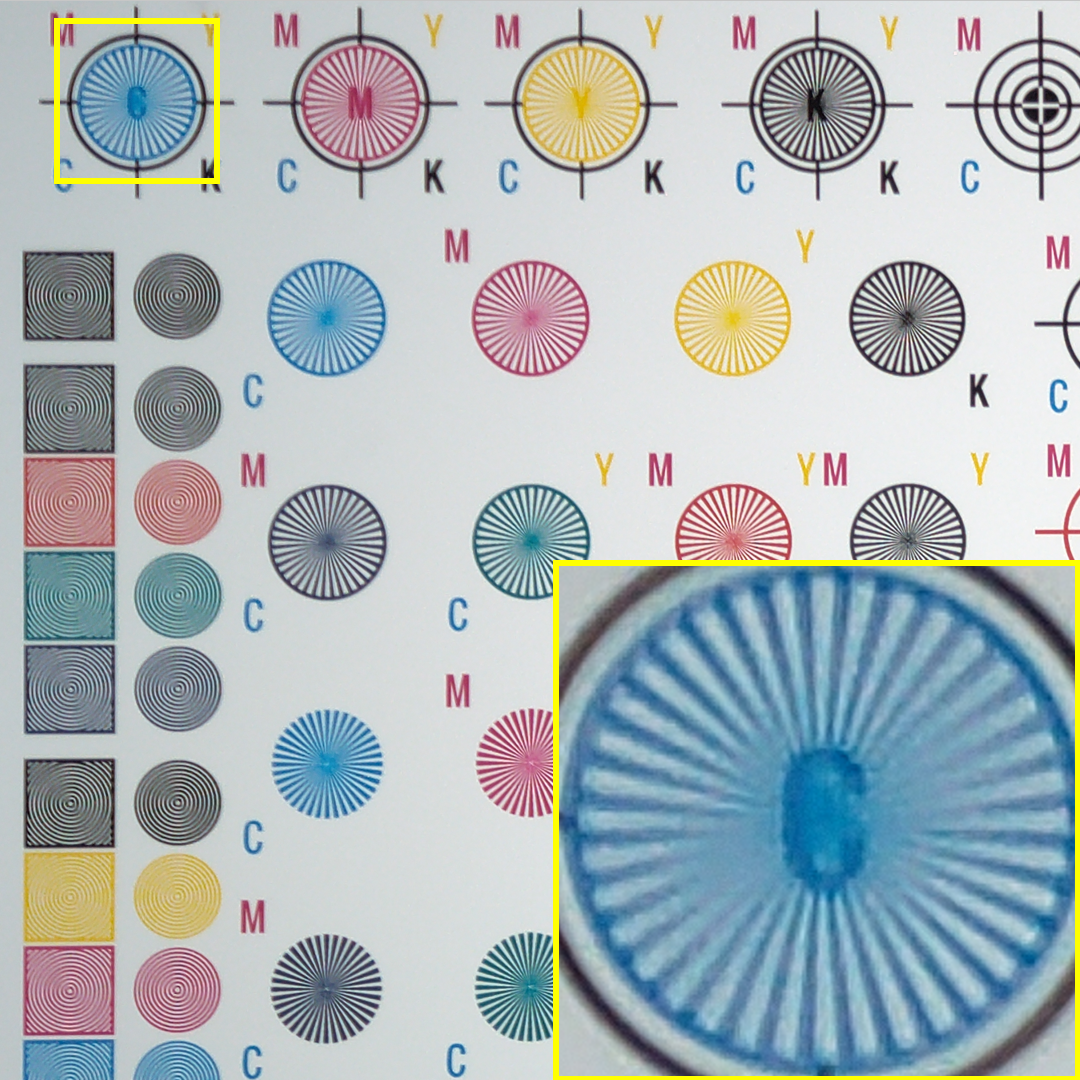}
\label{fig:conventional}
\end{subfigure}
\hfill
\begin{subfigure}{0.24\textwidth}
\includegraphics[width=0.95\textwidth]{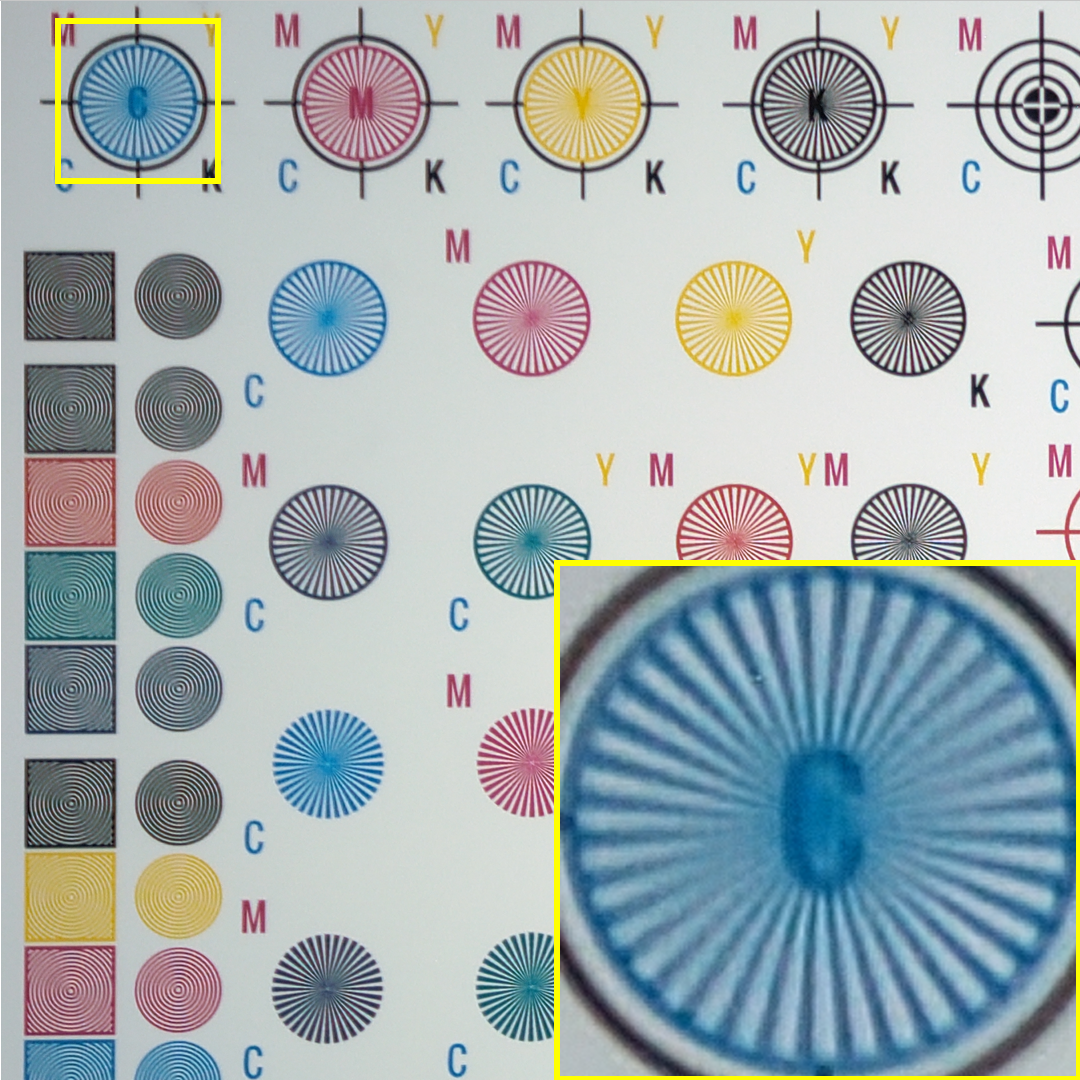}
\label{fig:pynet_high}
\end{subfigure}
\hfill
\begin{subfigure}{0.24\textwidth}
\includegraphics[width=0.95\textwidth]{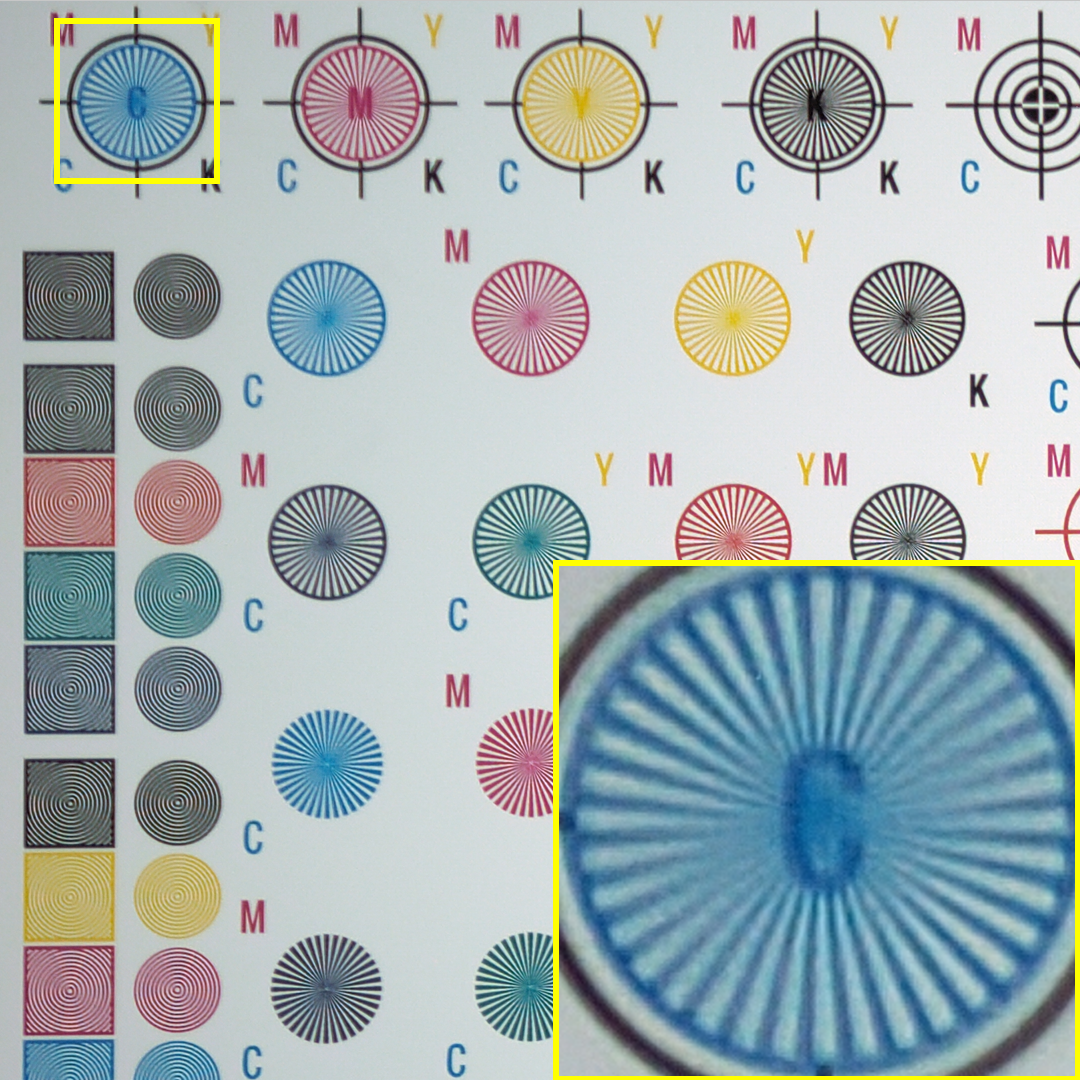}
\label{fig:enhen_high}
\end{subfigure}
\hfill
\begin{subfigure}{0.24\textwidth}
\includegraphics[width=0.95\textwidth]{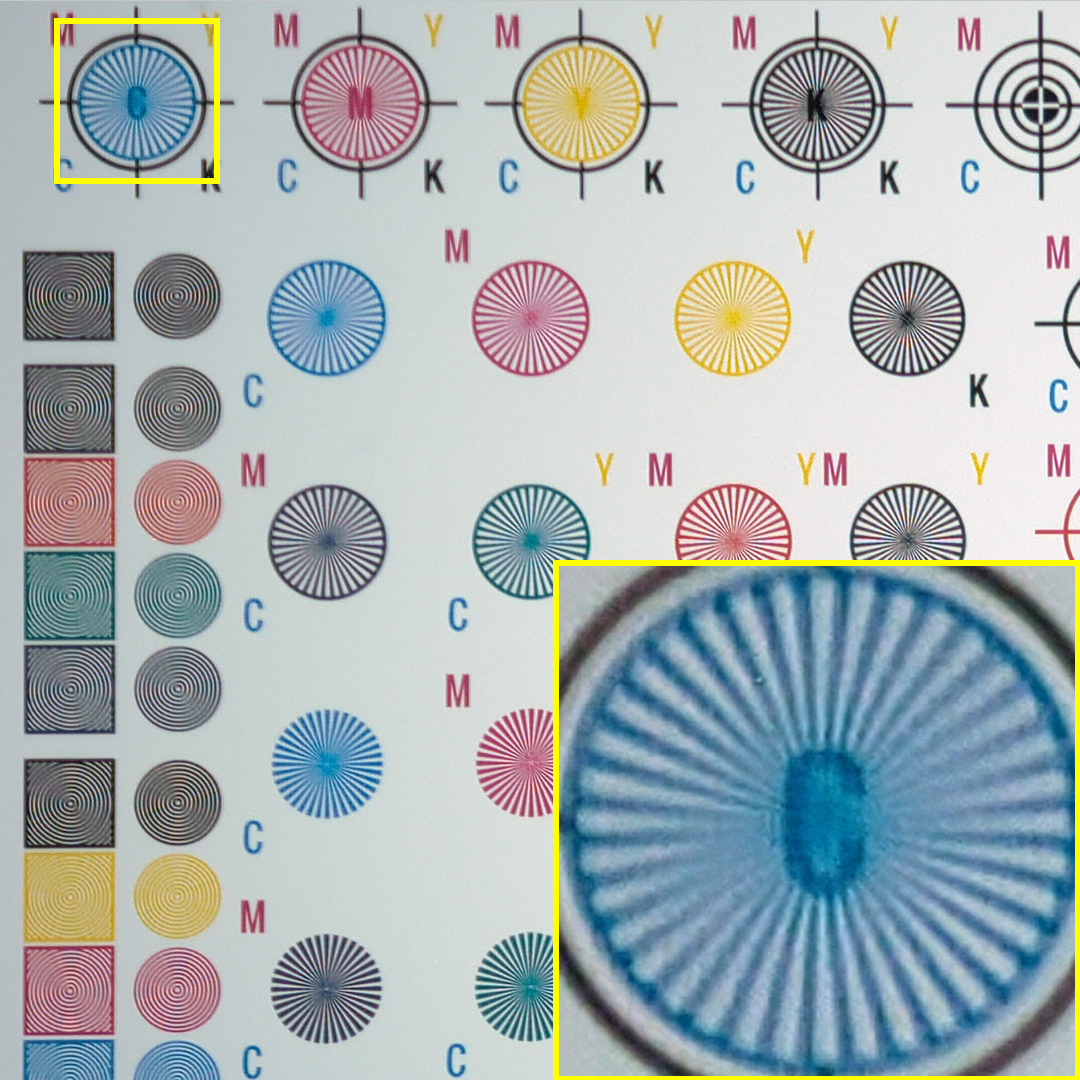}
\label{fig:ad_high}
\end{subfigure}
\caption{Visual comparison between conventional algorithm and deep learning algorithm for Q$\times$Q  demosaicing.
        From left to right, conventional, PyNET, enhanced PyNET, and PyNET-Q$\times$Q  with progressive distillation (Ours).}
\label{fig:conv_vs_dl}
\end{figure*}

\begin{figure*}
\centering
\begin{subfigure}{0.19\textwidth}
\includegraphics[width=0.95\textwidth]{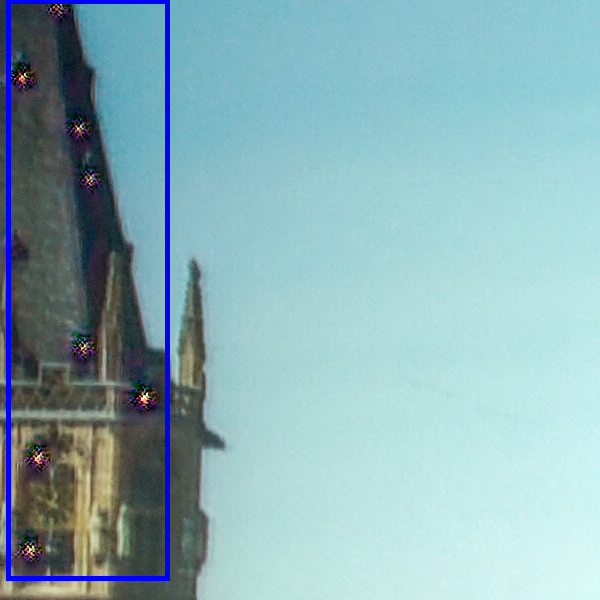}
\end{subfigure}
\begin{subfigure}{0.19\textwidth}
\includegraphics[width=0.95\textwidth]{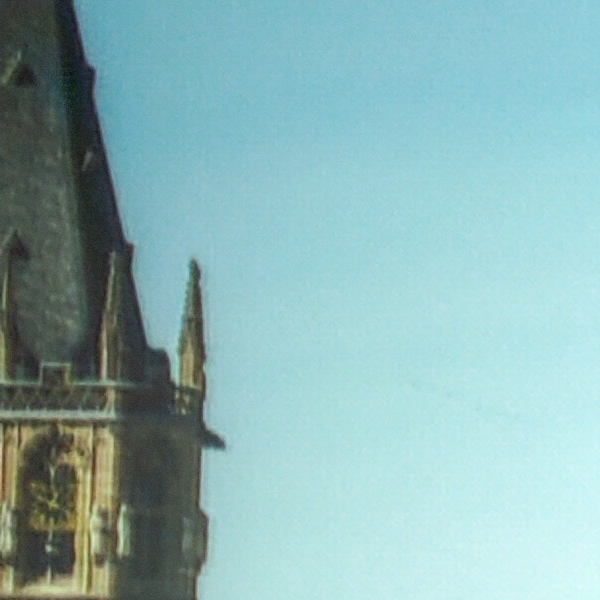}
\end{subfigure}
\begin{subfigure}{0.19\textwidth}
\includegraphics[width=0.95\textwidth]{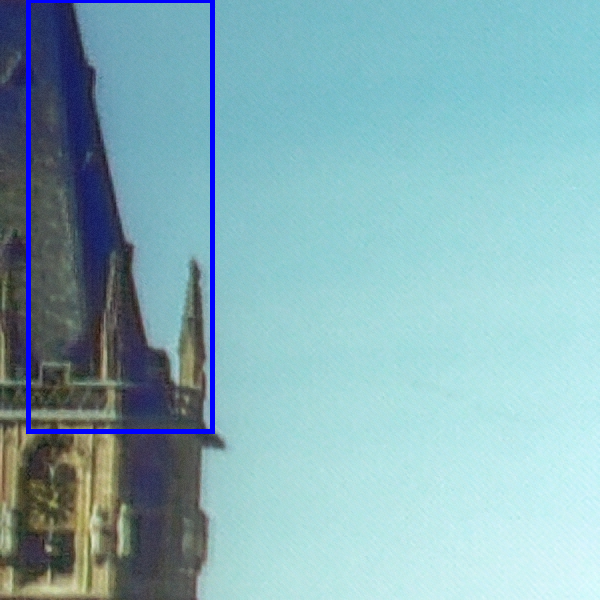}
\end{subfigure}
\begin{subfigure}{0.19\textwidth}
\includegraphics[width=0.95\textwidth]{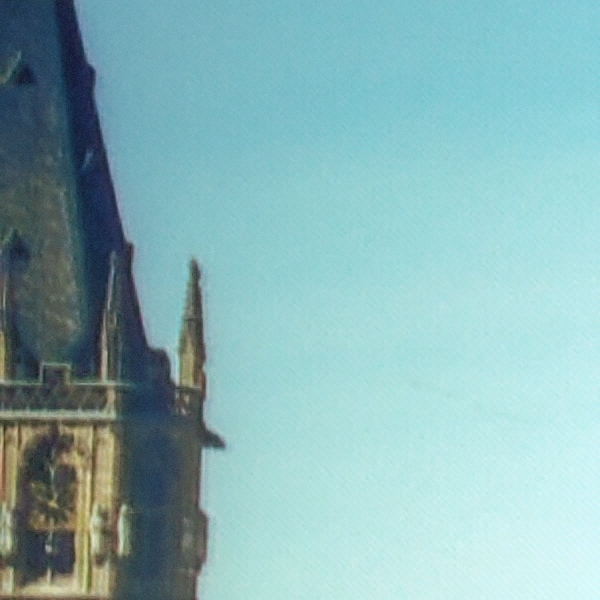}
\end{subfigure}
\begin{subfigure}{0.19\textwidth}
\includegraphics[width=0.95\textwidth]{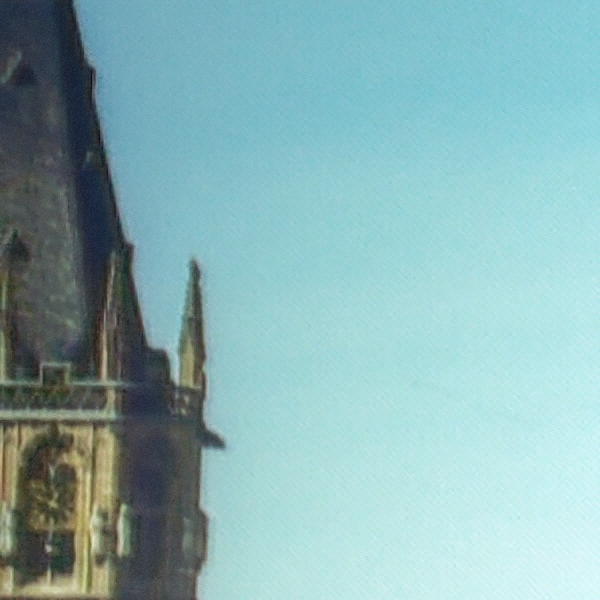}
\end{subfigure}\\
\begin{subfigure}{0.19\textwidth}
\includegraphics[width=0.95\textwidth]{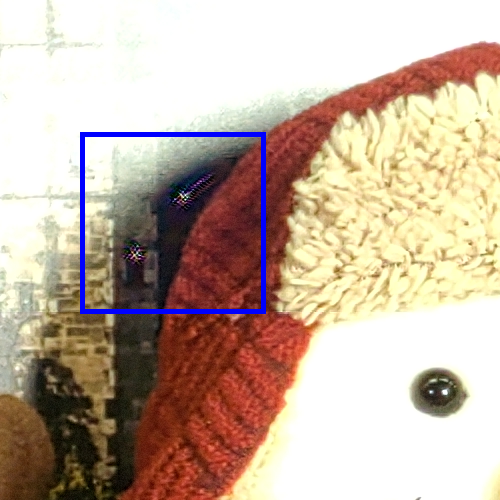}
\label{fig:pynet_res}
\end{subfigure}
\begin{subfigure}{0.19\textwidth}
\includegraphics[width=0.95\textwidth]{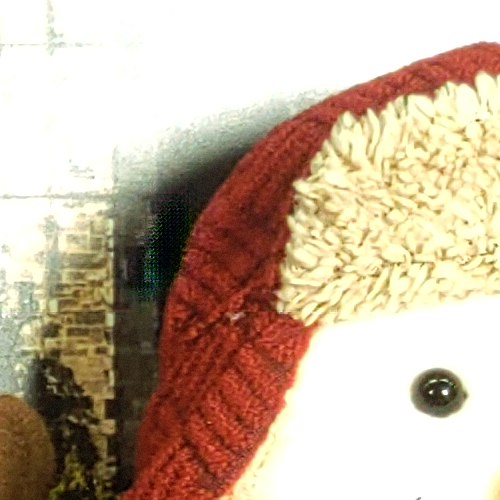}
\label{fig:enpynet_res}
\end{subfigure}
\begin{subfigure}{0.19\textwidth}
\includegraphics[width=0.95\textwidth]{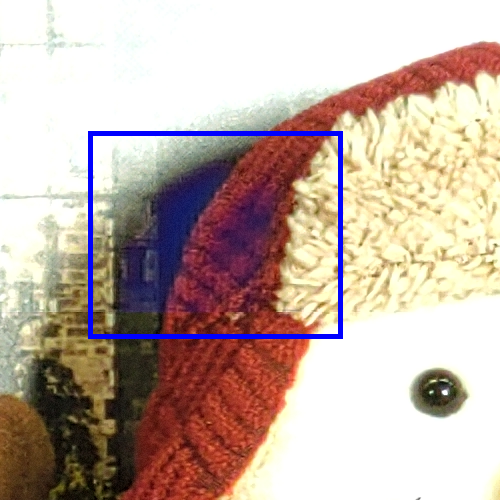}
\label{fig:pynetqq_res}
\end{subfigure}
\begin{subfigure}{0.19\textwidth}
\includegraphics[width=0.95\textwidth]{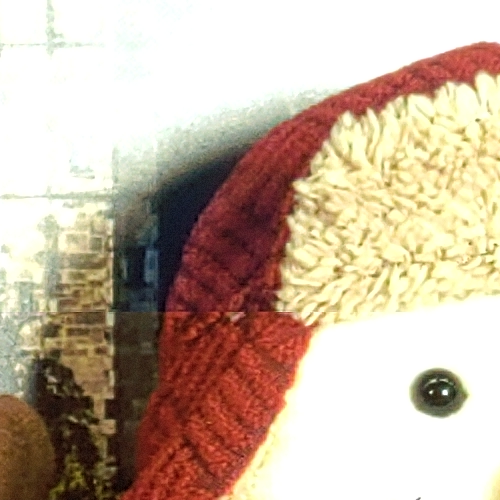}
\label{fig:kd_res}
\end{subfigure}
\begin{subfigure}{0.19\textwidth}
\includegraphics[width=0.95\textwidth]{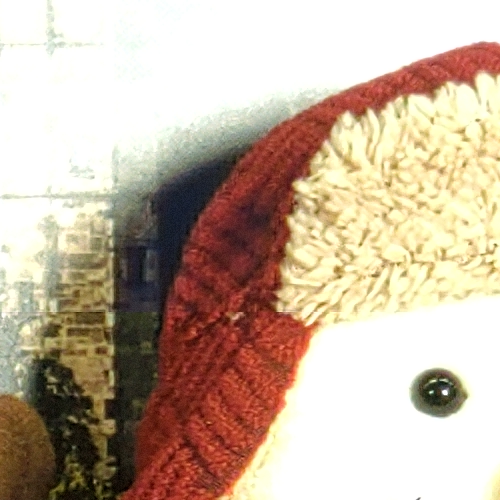}
\label{fig:ad_res}
\end{subfigure}
\caption{Visual comparison of PyNET based models.
From left to right: PyNET, Enhanced PyNET, PyNET-Q$\times$Q , PyNET-Q$\times$Q  with knowledge distillation,
and PyNET-Q$\times$Q  with progressive distillation (Ours).}
\label{fig:pynetcompare}
\end{figure*}

\begin{figure*}[!t]
     \centering
     \begin{subfigure}{0.23\textwidth}
    \includegraphics[width=0.95\textwidth]{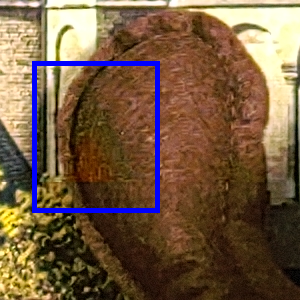}
    \caption{Base}
    \label{fig:baseline_object}
    \end{subfigure}
\hfill
    \begin{subfigure}{0.23\textwidth}
    \includegraphics[width=0.95\textwidth]{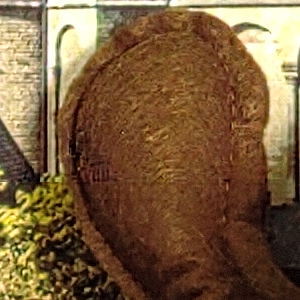}
    \caption{Base$+$SC}
    \label{fig:withgray_object}
    \end{subfigure}
\hfill
    \begin{subfigure}{0.23\textwidth}
    \includegraphics[width=0.95\textwidth]{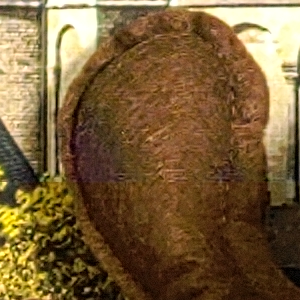}
    \caption{Base$+$SP}
    \label{fig:withsp_object}
    \end{subfigure}
\hfill
    \begin{subfigure}{0.23\textwidth}
    \includegraphics[width=0.95\textwidth]{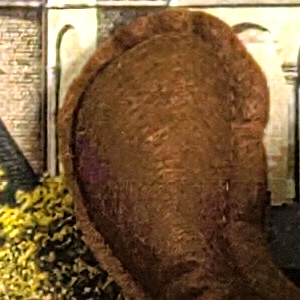}
    \caption{Base$+$SC$+$SP}
    \label{fig:withboth_object}
    \end{subfigure}
\caption{Ablation study on enhancement of PyNET-Q$\times$Q. (SC: skip connection with a gray image,
SP: sub-pixel convolution).}
\label{fig:albation}
\end{figure*}

\begin{figure*}[!h]
\centering
\subfloat[DIV2K]{\includegraphics[width=0.31\textwidth]{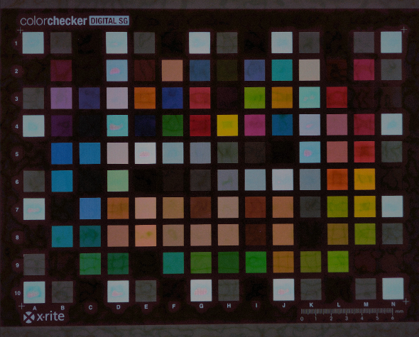}}
\label{fig:div2k}
\subfloat[3CCD]{\includegraphics[width=0.31\textwidth]{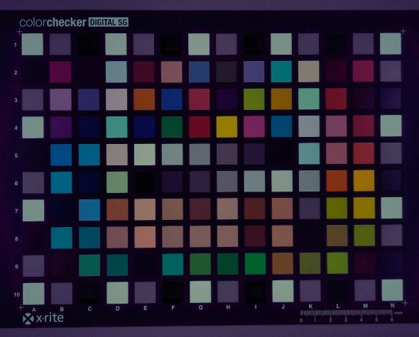}}
\label{fig:3ccd}
\subfloat[hybrid]{\includegraphics[width=0.31\textwidth]{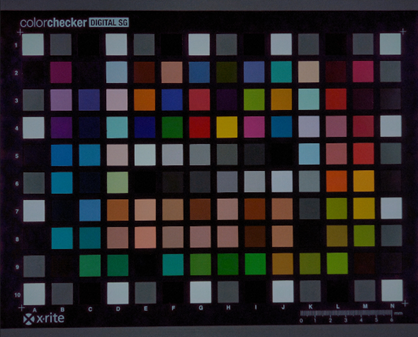}}
\label{fig:hybrid}
\caption{Visual comparison of trained PyNET with different datasets.}
\label{fig:dataset}
\end{figure*}

In this work, we incorporate RAW 3CCD images (captured by Hitachi HV-F203SCL) into our dataset.
A 3CCD camera splits incoming light into three RGB color beams using a dichroic prism,
and three color sensors measure each beam's intensity.
As a result, it can obtain RGB RAW images at the sensor level without an additional ISP system.
There is no loss of original color information or positional mismatch in 3CCD RAW images.
We train PyNET-Q$\times$Q on a {\it hybrid dataset}, combining 935 RAW 3CCD images and 900 images 
from the DIV2K dataset~\cite{div2k} for training and validation.
Since common datasets contain ISP-processed images, we apply an inverse gamma function with $\gamma=2.2$ 
(a common choice for balancing monitors and true color) to match input statistics by reverting gamma correction.
We demonstrate that the hybrid dataset enhances the model's capability in experiments,
with visual comparisons provided in Section~\ref{subsubsec:dataset}. 
Following PyNET~\cite{pynet}, processed images are cropped into $448\times448$ size patches.
Patches with low pixel variance are discarded for training stability.
The hybrid dataset comprises 14,595 training patches and 481 test patches.
We evaluate the trained model using Q$\times$Q images acquired from an actual Q$\times$Q image sensor
(under development). 
Due to their large size ($8000 \times 6000$), the test Q$\times$Q RAW images are cropped into $2912\times2912$ patches.
Sample 3CCD images and Q$\times$Q images with detailed descriptions are provided in Appendices.
Both RAW 3CCD data and Q$\times$Q test RAW images are 10-bit images, offering more detailed information than 8-bit images.

\begin{table*}[!ht]
\setlength{\tabcolsep}{4pt}
\begin{center}
\caption{PSNR, MS-SSIM, and FLOPs of test images ($448\times 448$).
The {\bf bold} and \textcolor{blue}{blue} text indicate the best and second-best scores, respectively.
(KD: knowledge distillation, PD: progressive distillation)}
\label{table:kd}
\begin{tabular}{lcccc}
\hline\noalign{\smallskip}
Methods & PSNR & MS-SSIM & Paramerters (M) & FLOPS (MAC) \\
\noalign{\smallskip}
\hline
PyNET                     & 34.964    & \textcolor{blue}{0.9841}    & 47.55 M & 342.35 G \\
Enhanced PyNET (Teacher)  & {\bf 36.650}    &{\bf 0.9870}    & 56.97 M & 342.73 G \\
PyNET-Q$\times$Q  (Student)    & 34.564    & 0.9825    & 1.07 M  & 54.05 G \\
PyNET-Q$\times$Q  $+$ KD       & 34.562    & 0.9813    & 1.07 M  & 54.05 G \\
PyNET-Q$\times$Q  $+$ PD (Ours)& \textcolor{blue}{34.973}    & 0.9833    & 1.07 M  & 54.05 G \\
\noalign{\smallskip}
\hline
\end{tabular}
\end{center}
\end{table*}
\setlength{\tabcolsep}{1.4pt}

\setlength{\tabcolsep}{4pt}
\begin{table}[!h]
\begin{center}
\caption{Ablation study.
The baseline is the PyNET with two levels (level 0 and 1) and half of the filters. 
(SC: skip connection with a gray image, SP: sub-pixel convolution)}
\label{table:ablation}
\begin{tabular}{lcccc}
\hline\noalign{\smallskip}
Methods & SC & SP & PSNR & MS-SSIM \\
\noalign{\smallskip}
\hline
baseline                &           &           &32.594 &0.9754 \\
baseline$+$SC           &\checkmark &           &33.787 &0.9800 \\
baseline$+$SP           &           &\checkmark &33.972 &0.9803 \\
baseline$+$SC$+$SP(Ours)&\checkmark &\checkmark &34.564 &0.9825 \\
\noalign{\smallskip}
\hline
\end{tabular}
\end{center}
\end{table}
\setlength{\tabcolsep}{1.4pt}

\subsubsection{3CCD Dataset}\label{app:3ccd}
We supply 100 3CCD images ($1600\times 1200$), with 50 indoor and 50 outdoor images.
The images are in `.RAW' format, where all pixel information is cascaded in a single line.
The RAW file first contains red values of all pixels, followed by green and blue values.
Each color information is described in 10 bits and stored using 2 bytes.
Consequently, a 3CCD RAW file contains $width \times height \times 3 \times 2$ bytes. 
Additionally, the file includes a global black level offset of 64,
requiring a black level compensation of 64 from all pixel color values.

\subsubsection{Q$\times$Q Dataset}\label{app:qxq}
We offer six sample Q$\times$Q images ($8000 \times 6000$) obtained from an actual Q$\times$Q camera sensor
(under development).
Since these Q$\times$Q images are sensor-level images,
they possess different characteristics compared to other common images.
For example, pixel values of green have a bias even in plain white regions,
resulting in a green color bias in the demosaiced image.
Later ISP steps will adjust this color bias.
Since each pixel in the Q$\times$Q image contains single color information, a single-channel image is stored.
Similar to 3CCD RAW files, all pixel values are cascaded in a single line in the Q$\times$Q RAW file.
Each color information is described in 10 bits and stored using 2 bytes.
Thus, a Q$\times$Q RAW file contains $width \times height \times 2$ bytes.
The Q$\times$Q RAW file also has a black level offset of 64, necessitating black level compensation.

%%% Results
%%%%%%%%%%%%%%%%%%%%%%%%%%%%%%%%%%%%%%%%%%%%%%%%%%%%%%%%%%%%%%%%%%%%
\subsection{Results}\label{subsec:results}
%%% Comparison with Other Methods
\subsubsection{Comparison with other methods}

As this is the first demosaicing model for Q$\times$Q images, 
we compare PyNET-Q$\times$Q with other PyNET variants (PyNET and enhanced) and conventional logic.
Throughout the experiment, we train all models (proposed and for comparisons) using a hybrid dataset.
We assess the models on actual Q$\times$Q images by visually examining reconstructed outputs,
as shown in Figure~\ref{fig:conv_vs_dl}\footnote{Q$\times$Q  images cannot be evaluated quantitatively 
because there are no ground truth images.}.
The conventional method lacks smooth edge reconstruction, while PyNET-based models exhibit smooth edges,
as depicted in the upper figures.
Moreover, the conventional method tends to intensify pixels of different colors in the RAW Q$\times$Q image,
as shown in the blue box in the upper figure.
The high-frequency reconstruction of the conventional method is somewhat blurry,
while PyNET-based models recreate more defined shapes, as demonstrated in the lower figures.
Surprisingly, PyNET-Q$\times$Q displays visual quality comparable to enhanced PyNET even with a 1/50 parameter reduction.
Additional visual comparisons between the conventional method and PyNET-Q$\times$Q can be found in Appendices.

\subsubsection{Impact of progressive distillation}
We investigate the effectiveness of progressive distillation.
In Table~\ref{table:kd}, the enhanced PyNET outperforms other PyNET variants in terms of PSNR and MS-SSIM.
Despite having fewer parameters, PyNET-Q$\times$Q with progressive distillation (PD) achieves 
the second-highest PSNR score, even surpassing the original PyNET.
Notably, PyNET-Q$\times$Q without distillation presents lower scores than the original PyNET,
indicating that our enhancement and progressive distillation successfully compensate for level removal.
Besides the scores, PyNET-Q$\times$Q without distillation exhibits a lower visual quality of reconstruction,
particularly in texture.
For instance, it contains a blue-ish shadow in Figure~\ref{fig:pynetcompare}.
This is because PyNET-Q$\times$Q without distillation struggles to handle global texture and colors,
roles played by lower levels in the original PyNET.
However, progressive distillation transfers the `knowledge' of lower levels from enhanced PyNET to PyNET-Q$\times$Q,
enabling more realistic reconstructions.

Compared to the original PyNET, which has the second-best MS-SSIM,
the proposed model with progressive distillation exhibits comparable MS-SSIM.
However, the original PyNET reveals some artifacts in shadows, as shown in Figure~\ref{fig:pynetcompare}.
On the other hand, PyNET-Q$\times$Q with progressive distillation improves these artifacts.
More visual comparisons can be found in the Supplementary Materials.

%%% Ablation Study
\subsubsection{Ablation study}
To confirm the effectiveness of progressive distillation,
we conducted experiments with PyNET-Q$\times$Q (a) without distillation, (b) with regular knowledge distillation,
and (c) with progressive distillation, as shown in Table~\ref{table:kd}.
To evaluate the additional components of enhanced PyNET, we performed the following supplementary experiments.
For a fair comparison, the number of model parameters is the same across all experiments.
Since the aim of the proposed demosaicing algorithm is to make PyNET lighter and enhance performance,
we use PyNET with two levels (level~0 and~1) and half of the filters as our baseline.

The proposed enhancements are 1) a skip connection with a gray image (SC) and 2) sub-pixel convolution (SP).
The scores (PSNR and MS-SSIM) of baseline models with and without 
each enhancement are presented in Table~\ref{table:ablation}.
The skip connection with a gray image improves PSNR by 1.2 dB compared to the baseline.
It also exhibits a better MS-SSIM result, 
confirming that the skip connection with a gray image enhances perceptual image quality.
The visual evaluation also demonstrates the contribution of the skip connection with a gray image,
as depicted in Figure~\ref{fig:baseline_object} and Figure~\ref{fig:withgray_object}.
The baseline exhibits a red false color artifact and distorted tone and texture.
However, the model with a skip connection with a gray image preserves the original color 
and more accurately restores the original texture.

The sub-pixel convolution upsampling increases PSNR by 1.38 dB compared to the baseline.
The significant improvement in PSNR and MS-SSIM highlights the importance of the appropriate upsampling method.
The sub-pixel convolution upsampling also yields better color reconstruction,
while the baseline displays color deviation, as illustrated in Figure~\ref{fig:baseline_object} 
and Figure~\ref{fig:withsp_object}.
The result suggests that sub-pixel convolution upsampling outperforms interpolation-based upsampling.
By incorporating both skip connection with a gray image and subpixel-based upsampling,
the model substantially improves PSNR, MS-SSIM, and visual quality, as shown in Figure~\ref{fig:withboth_object}.

%%% dataset compare
\subsubsection{Impact of 3CCD and Hybrid Datasets} \label{subsubsec:dataset}
We investigate the influence of datasets on demosaicing model training.
We train PyNET models using a 3CCD RAW dataset, DIV2K dataset, and hybrid dataset, respectively.
The 3CCD dataset consists of 5,827 images ($400\times400$) cropped from 528 RAW 3CCD images ($1600 \times 1200$),
whereas the DIV2K dataset is comprised of 10,715 images cropped from 800 DIV2K HR images,
and the hybrid dataset utilizes both DIV2k and 3CCD data.
We maintain the same number of epochs, so the learning amount differs due to varying dataset sizes.
Output images generated by these models are displayed in Figure~\ref{fig:dataset}.
The model trained with the DIV2K dataset exhibits color mismatches in monochromatic regions.
In contrast, the model trained with the 3CCD dataset demonstrates better color restoration 
in low-light environments but has a color bias where gray blocks exhibit a purple tone.
Using the hybrid dataset, color restoration is further enhanced without color bias.

\section{Conclusions}
We introduced PyNET-Q$\times$Q, the first deep learning-based demosaicing model for Q$\times$Q images.
By incorporating a skip connection with a gray image and sub-pixel convolution,
we enhanced PyNET to better suit Q$\times$Q inputs.
We then compressed the enhanced PyNET and trained it using progressive distillation.
The proposed model is considerably smaller in size (2.3\% compared to PyNET)
while maintaining a comparable demosaicing capability.
We evaluated our model with input captured by an actual Q$\times$Q image sensor and demonstrated high-quality demosaiced images.

\section*{Acknowledgment}
The authors would like to thank Jonghyun Bae and Jinsu Kim for providing 3CCD and Q$\times$Q  images and supporting our experiments.

\clearpage

\bibliographystyle{splncs04}
\bibliography{reference}

\newpage
\appendix

\begin{figure*}[!h]
\centering
\subfloat[Q$\times$Q RAW input image]{\includegraphics[width=0.48\textwidth]{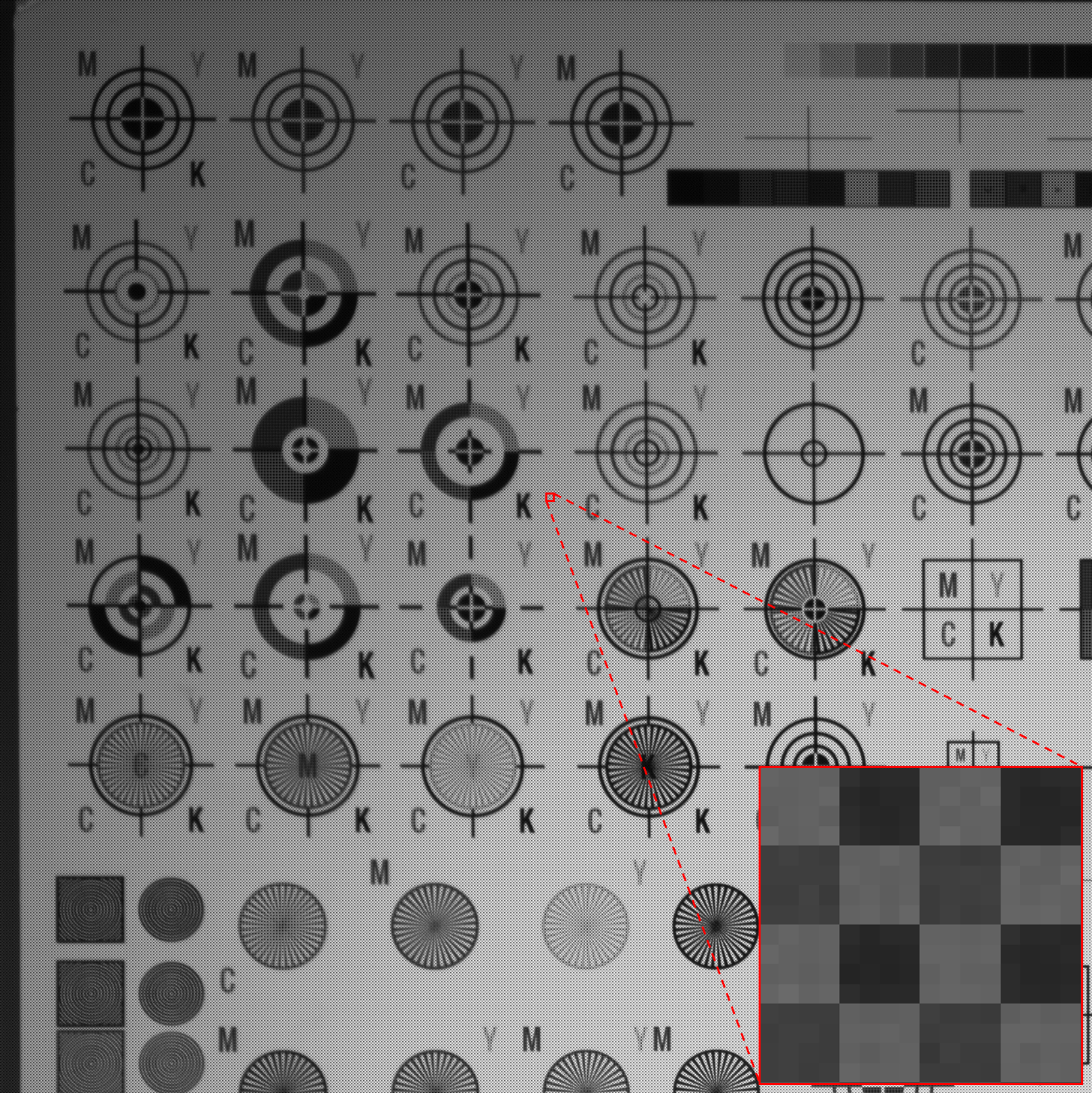}}
\label{fig:qxq raw green bias}
\subfloat[Demosaiced image]{\includegraphics[width=0.48\textwidth]{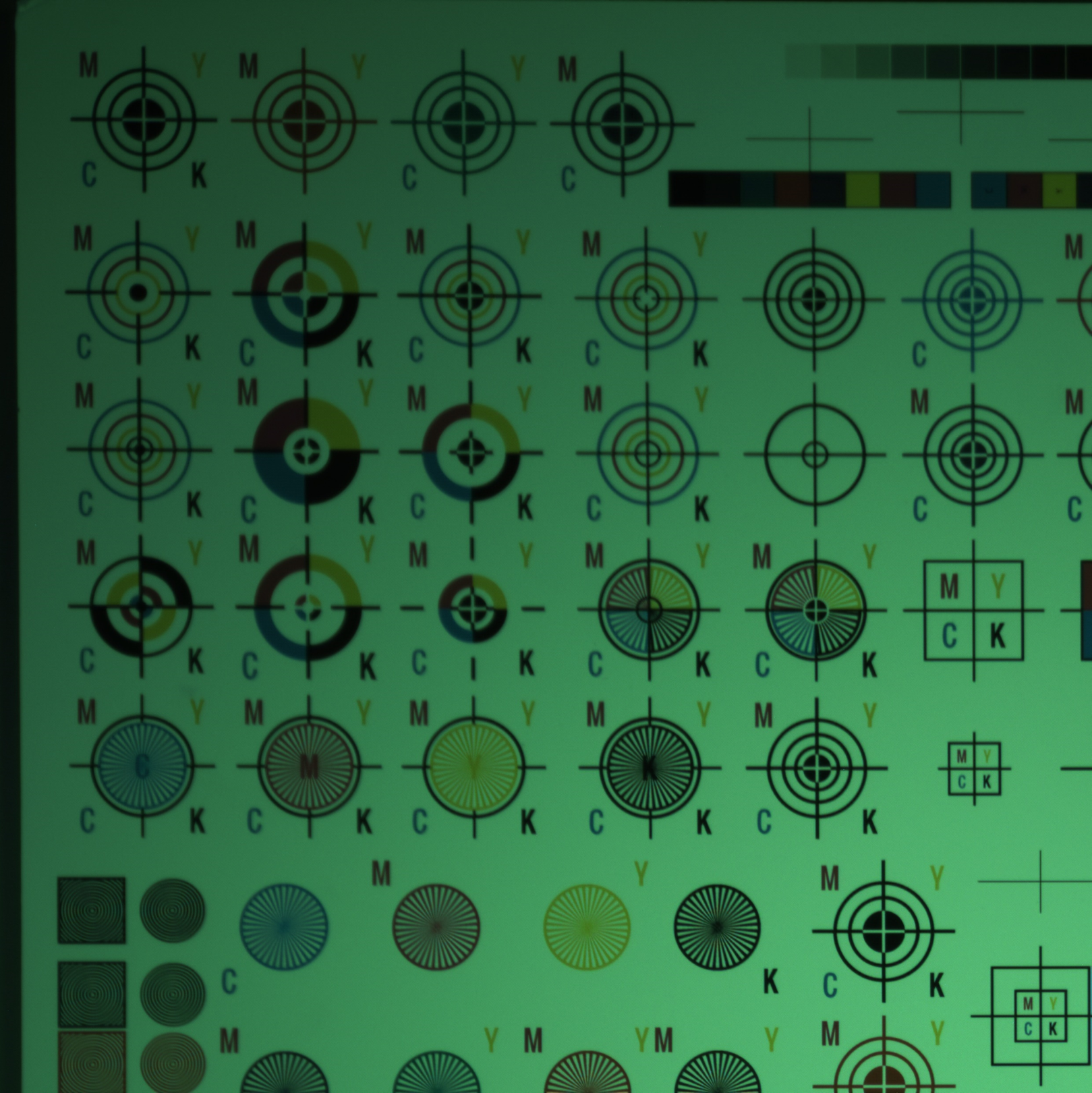}}
\label{fig:qxq demosaiced before isp}
\caption{Example of the Q$\times$Q RAW image and demosaiced image. The demosaiced image is before applying other ISP steps.}
\end{figure*}

%%% Comparison: conventional and deeplearning

\begin{figure*}[!h]
\centering
\subfloat[Conventional algorithm]{\includegraphics[width=0.47\textwidth]{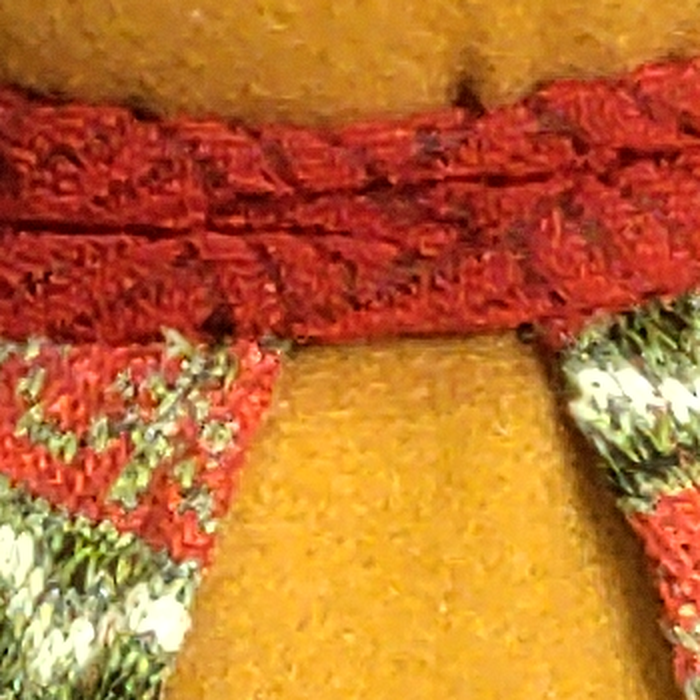}}
\label{fig:conv1}
\subfloat[Proposed algorithm]{\includegraphics[width=0.47\textwidth]{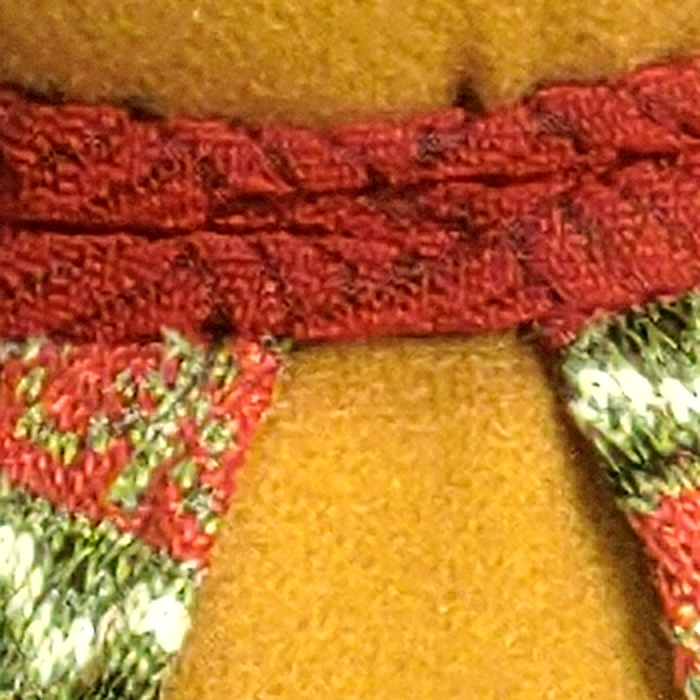}}
\label{fig:dl1}
\caption{Visual comparison of detailed texture reconstruction.}
\end{figure*}

\begin{figure*}[!h]
\centering
\subfloat[Conventional algorithm]{\includegraphics[width=0.49\textwidth]{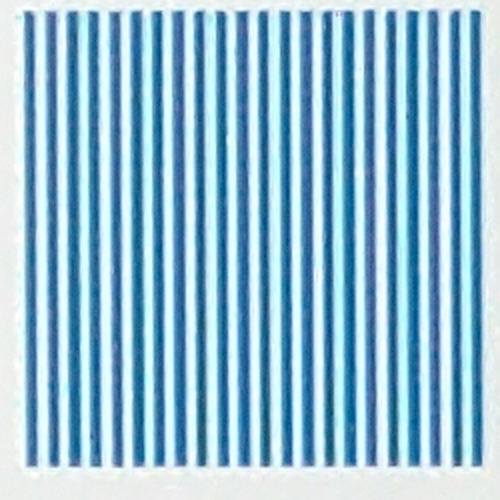}}
\hfill
\subfloat[Proposed algorithm]{\includegraphics[width=0.49\textwidth]{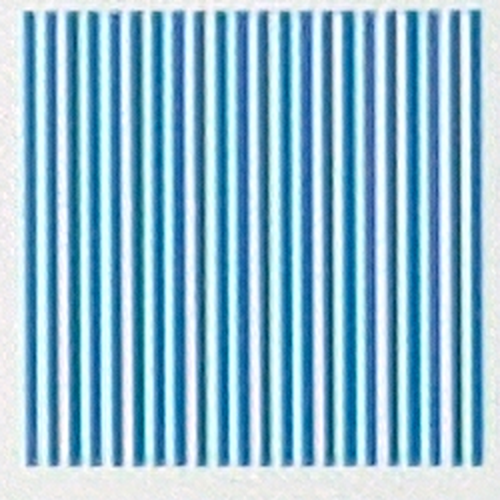}}
\caption{Visual comparison of reconstruction at high frequency}
\end{figure*}

\begin{figure*}[!h]
\centering
\subfloat[Conventional algorithm]{\includegraphics[width=0.49\textwidth]{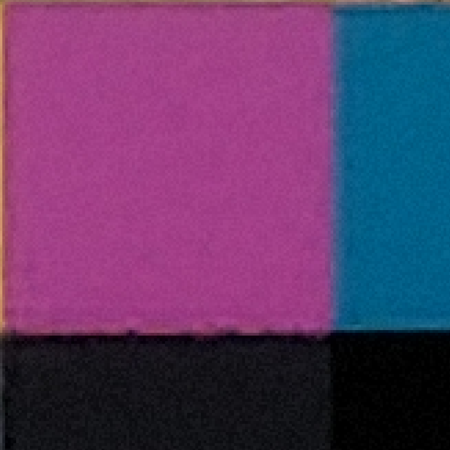}}
\label{fig:conv4}
\subfloat[Proposed algorithm]{\includegraphics[width=0.49\textwidth]{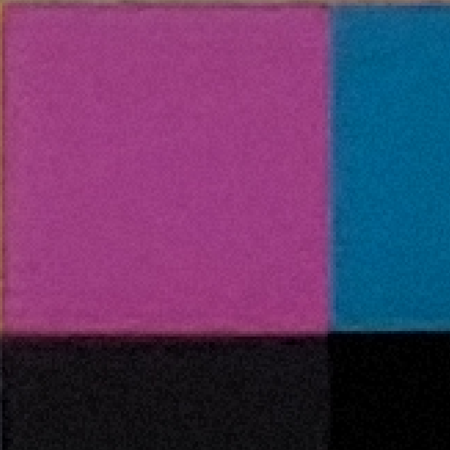}}
\label{fig:dl4}
\caption{Visual comparison of edge reconstruction}
\label{fig:convsdl4}
\end{figure*}

\begin{figure*}[!h]
\centering
\subfloat[Conventional algorithm]{\includegraphics[width=0.49\textwidth]{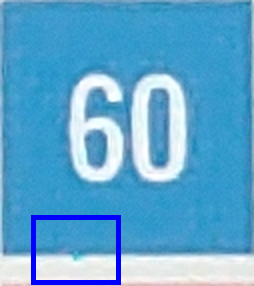}}
\subfloat[Proposed algorithm]{\includegraphics[width=0.49\textwidth]{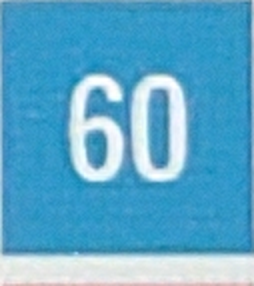}}
\\
\subfloat[Conventional algorithm]{\includegraphics[width=0.49\textwidth]{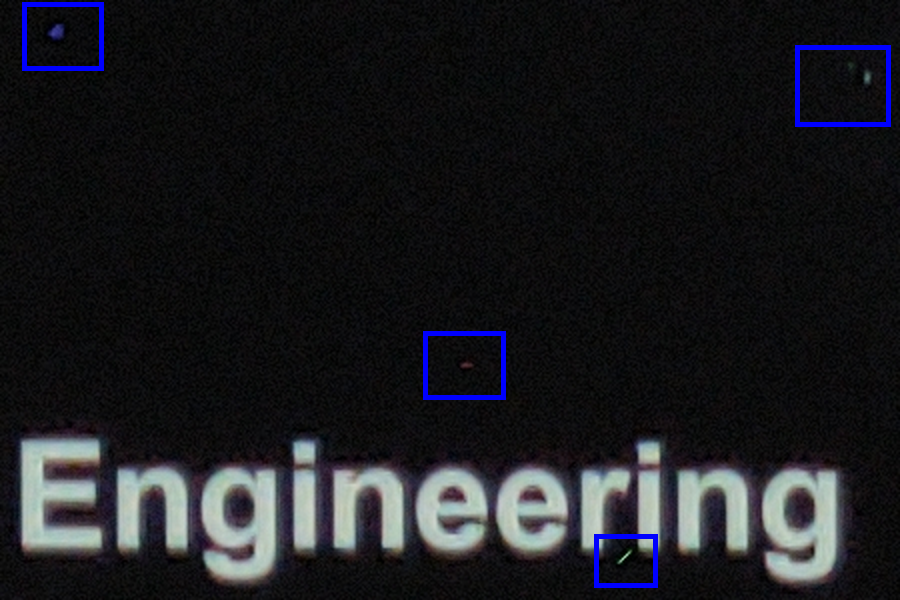}}
\label{fig:conv3}
\subfloat[Proposed algorithm]{\includegraphics[width=0.49\textwidth]{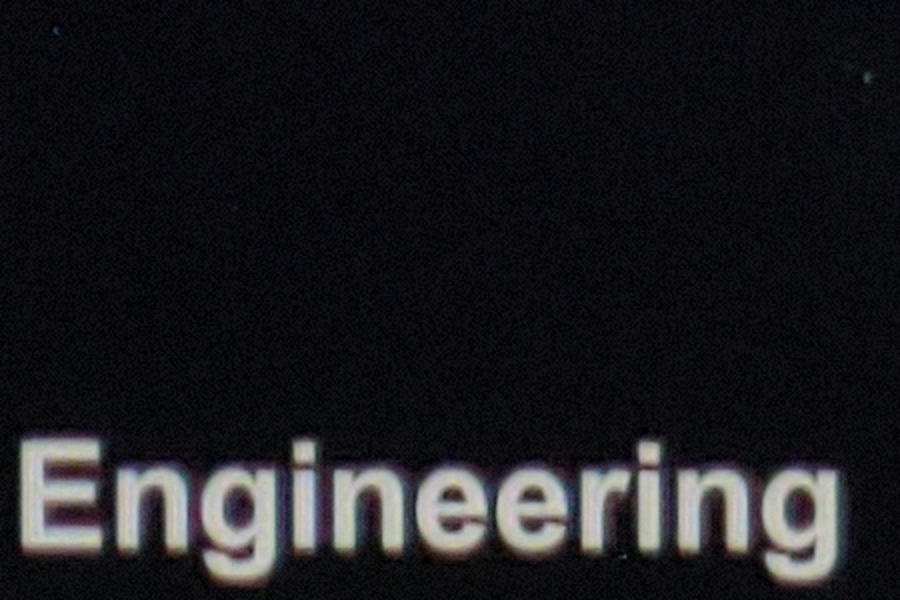}}
\label{fig:dl3}
\caption{Visual comparison of false color removal}
\end{figure*}

%%% pynet based 

\begin{figure*}[!h]
\centering
\subfloat[PyNET]{\includegraphics[width=0.33\textwidth]{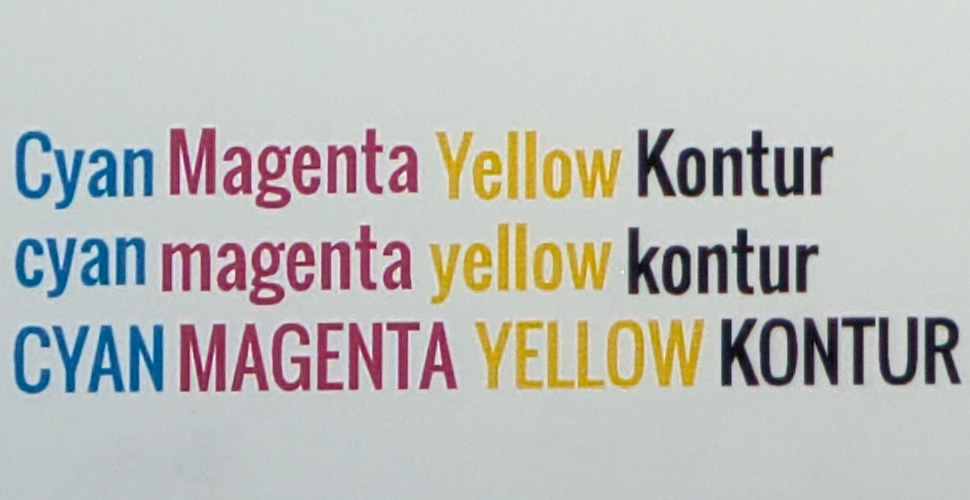}}
\hfill
\subfloat[Enhanced PyNET]{\includegraphics[width=0.33\textwidth]{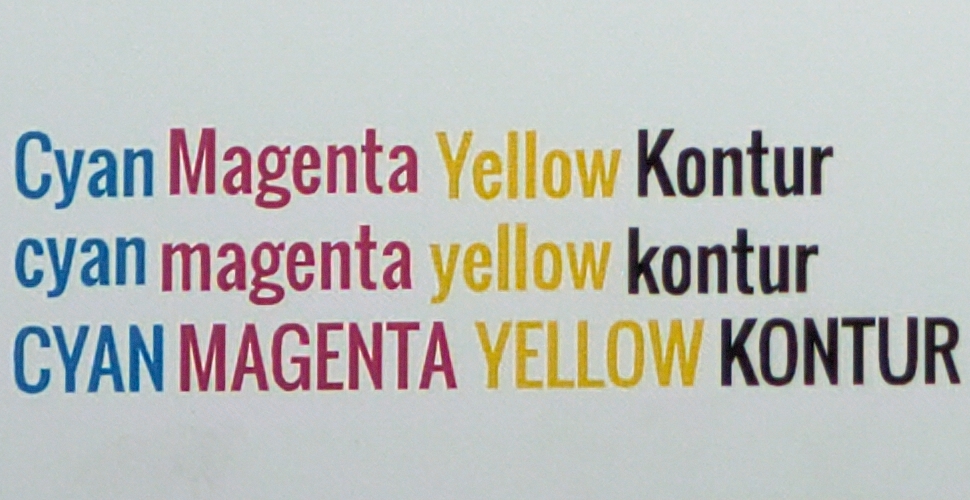}}
\hfill
\subfloat[Proposed algorithm]{\includegraphics[width=0.33\textwidth]{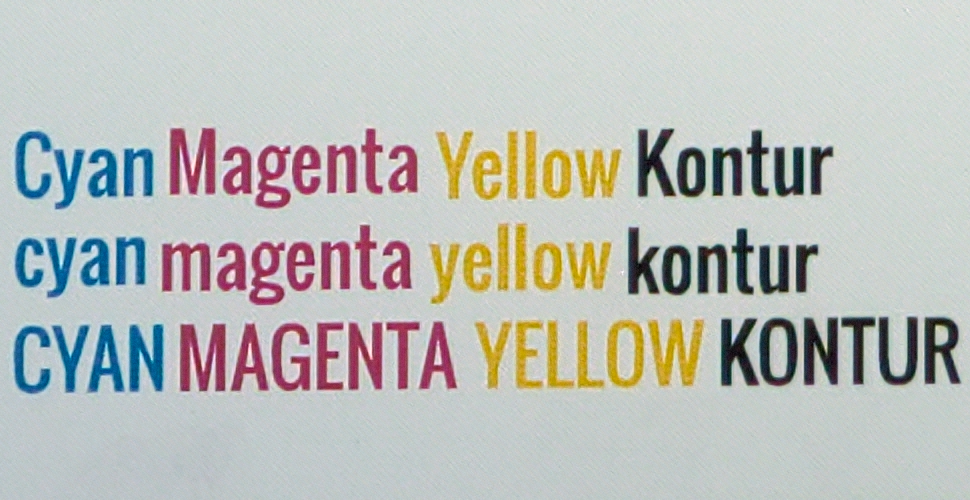}}
\caption{Visual comparison of text reconstruction}
\end{figure*}

\begin{figure*}[!h]
\centering
\subfloat[PyNET]{\includegraphics[width=0.33\textwidth]{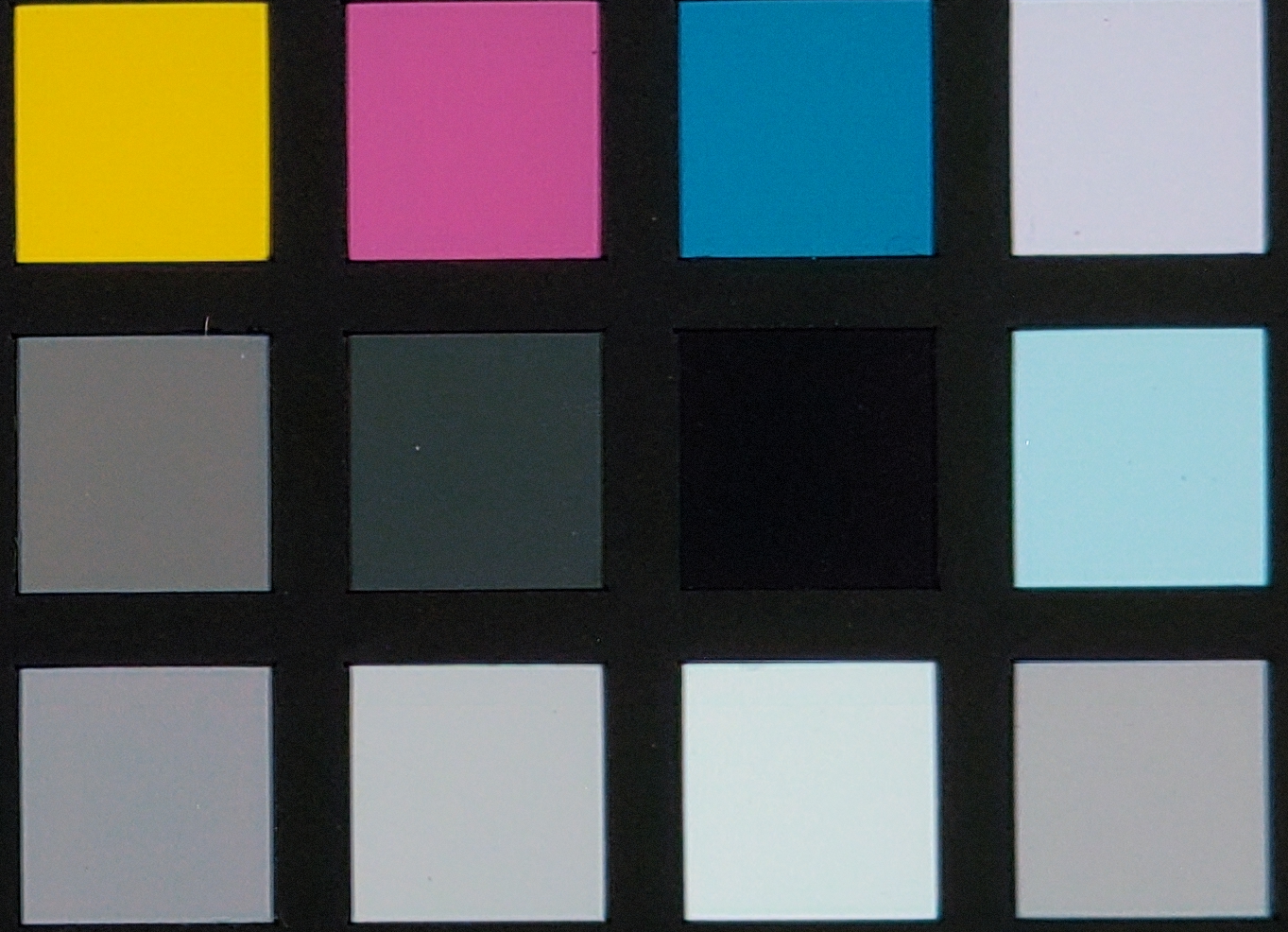}}
\hfill
\subfloat[Enhanced PyNET]{\includegraphics[width=0.33\textwidth]{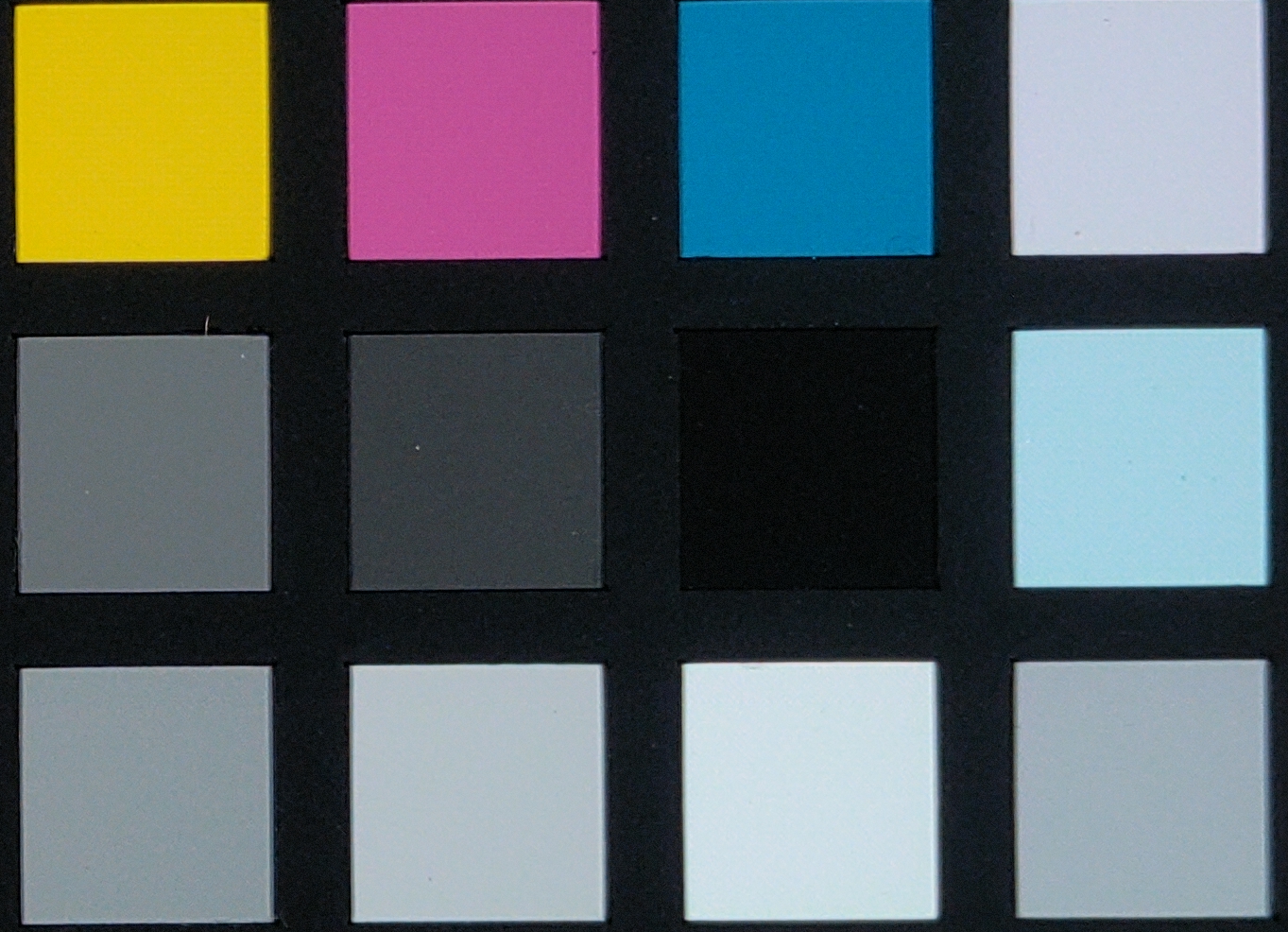}}
\hfill
\subfloat[Proposed algorithm]{\includegraphics[width=0.33\textwidth]{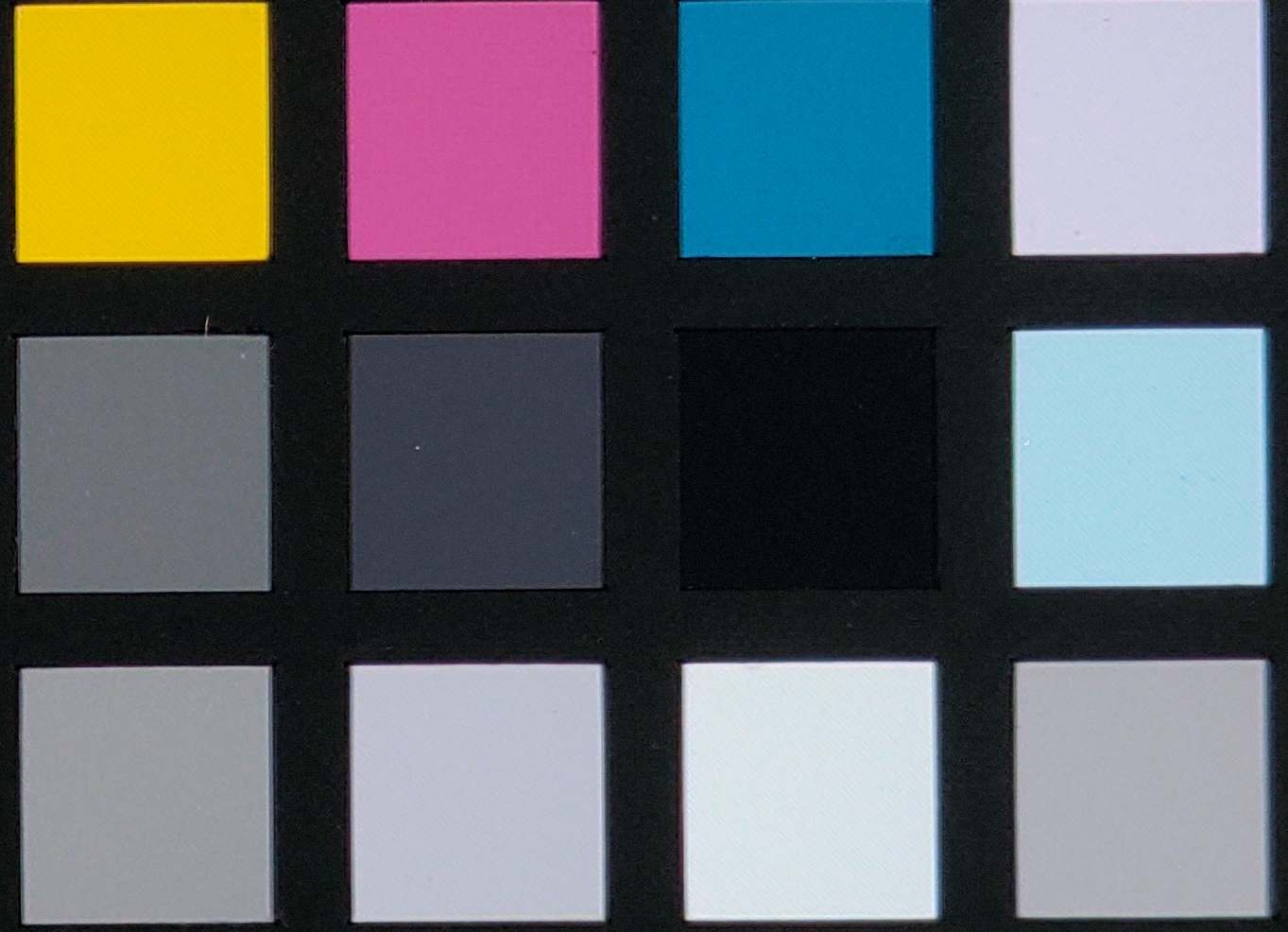}}
\caption{Visual comparison of edge reconstruction}
\end{figure*}

\begin{figure*}[!ht]
\centering
\subfloat[PyNET]{\includegraphics[width=0.33\textwidth]{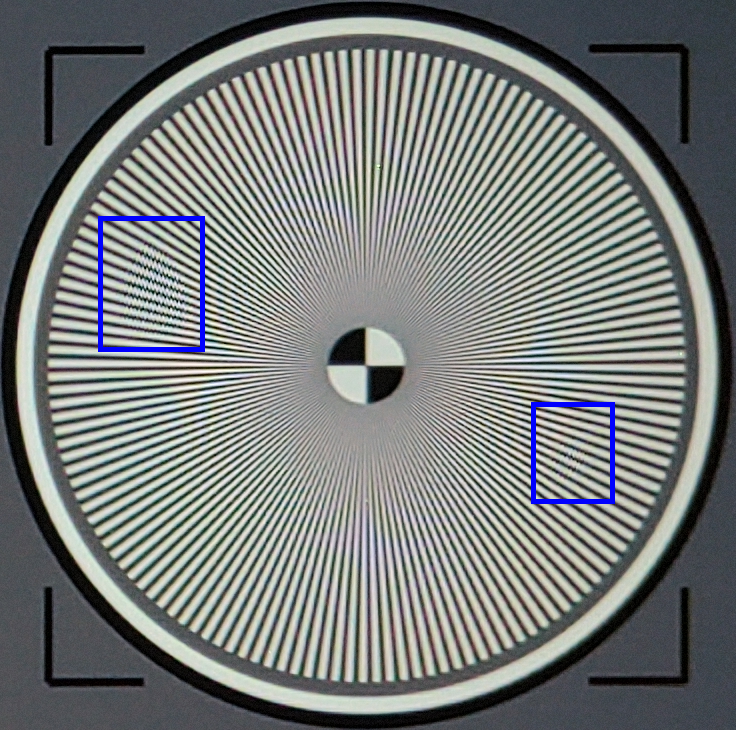}}
\hfill
\subfloat[Enhanced PyNET]{\includegraphics[width=0.33\textwidth]{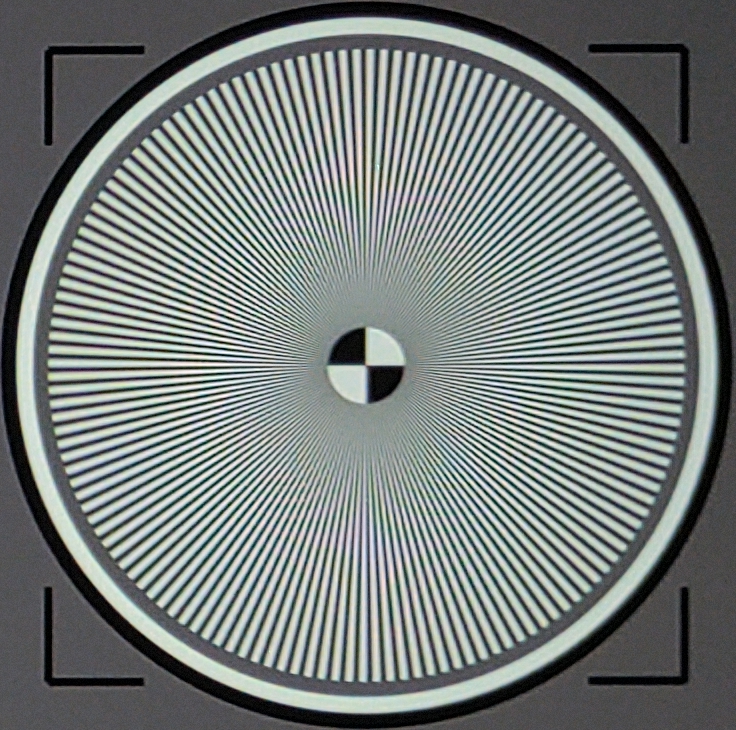}}
\hfill
\subfloat[Proposed algorithm]{\includegraphics[width=0.33\textwidth]{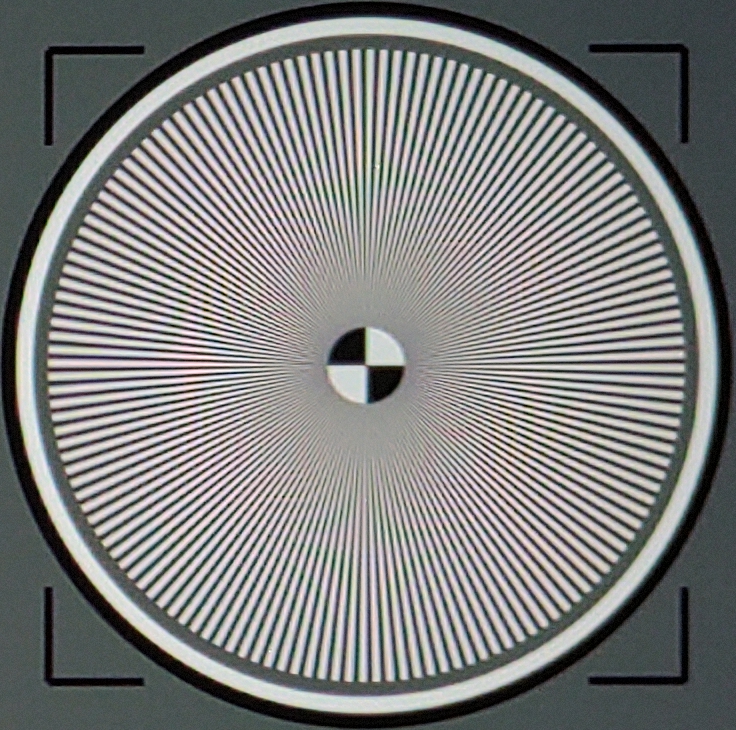}}
\caption{Visual comparison of reconstruction at high frequency} 
\end{figure*}

\end{document}